# MAGNETISM OF CARBON-BASED MATERIALS

**Tatiana Makarova**

*Ioffe Physico-Technical Institute, St. Petersburg, Russia,
Umeå University, Umeå, Sweden*
E-mail: tatiana.makarova@physics.umu.se

> For years there have been many reports of observations of weak spontaneous magnetization above room temperature in heat-treated organic compounds. In virtually every case, it has been difficult to establish the intrinsic character, or even the reproducibility, of the observed magnetism[12]. The possibility that some interesting magnetic behaviour might be going on behind those observations has never been addressed systematically.
>
> *F. Palacio, 2001*

## 1. Introduction

The history of organic magnets counts several decades. There has been recently a rapid progress in this area, and many high-temperature organic ferromagnets have been discovered in XXI century [1 - 4]. Starting with a brief description of molecular magnets having a determined chemical structure, the present paper shifts the focus to scattered reports on different metal-free organic compounds that exhibit ferromagnetic behavior even at room temperature. The carbon-based materials described in the majority of communications contain only a small part of organic ferromagnetic material, the results are technology-dependent and difficult to reproduce. The structural unit, giving rise to ferromagnetism in carbon-based structures, has not been yet found, and these materials can be called UFOs – Unidentified Ferromagnetic Organic compounds [2, p.124]. On the other hand, the analysis of numerous experimental results shows that they cannot be explained without recognition that ferromagnetic carbon does exist. This evidence supports the theoretical predictions showing that electronic instabilities in pure carbon may give rise to superconducting and ferromagnetic properties even at room temperature. Five types of carbon magnets have been obtained experimentally: (i) chains of interacting radicals (ii) carbonaceous substances with a mixture of sp$^2$ and sp$^3$ coordinated atoms; (iii) amorphous carbon structures containing trivalent elements like P, N, B; (iv) nanographite, nanodiamond, carbon nanofoam; (v) fullerenes.

## 2. Molecular magnets

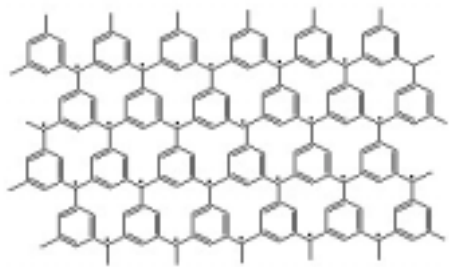

Fig. 1. A hypothetical high-spin two-dimensional structure (Mataga model).

The first predictions of organic magnetism are attributed to H. McConnell [5,6] and N. Mataga [7]. The earliest idea is now called the "McConnell-II model" and is interpreted in terms of the condition of orthogonality of magnetic orbitals for reaching a ferromagnetic exchange: spin exchange between positive spin density on one radical and negative spin density on another radical. The "McConnell-I model", extended by Breslow [8], deals with a special class of substances: quasi-one-dimensional spin chains .. D+•  A-•  D+•  A-• ... , where the cation-radicals of the donor D are alternated with the anion-radicals of the acceptor A. Based on the molecular orbital theory, N. Mataga, and also A. Ovchinnikov [9] proposed several π - conjugated polymers, such as polycarbenes and polyradicals, where the necessary conditions could be met: a spin alignment and a ferromagnetic coupling dominating throughout the solid. Fig. 1 shows an example of a planar two-dimensional structure, composed from triplet diphenylcarbenes integrated in an entire conjugate system through meta-positions of phenyl rings. The mechanism describing possible ferromagnetic states of similar hypothetical hydrocarbons is now called "Mataga model".

Since that time many materials with experimentally confirmed organic or molecular ferromagnetism have been obtained. Progress in this area first and foremost owes to J. S. Miller and A. J. Epstein. The first metal-organic substance [10] brought into existence was a series of tetracyanoethylenide – based molecular magnets with a general formula $M$ (TCNE)$_x$ • $y$ (solvent), where $M =$  V, Fe, Mn, Co, Ni. This family includes the first room-temperature organic magnet [11], air-stable thin-film plastic magnet with a Curie temperature of 370 K [12], and the first light-tunable organic magnet [13].

Another group of room-temperature molecular magnets is the Prussian Blue analogues. Ferrocyanide Prussian Blue $Fe^{III}_4 [Fe^{II} (CN)_6]_3 \cdot n\, H_2O$ undergoes ferromagnetic ordering at Tc = 5.6 K, but when the iron sites are substituted for other paramagnetic metal ions (V, Cr, Mn, Co and Ni) the Curie temperature increases to 315 K [14], 372 K [15], and above [16].

The first magnet composed of light elements, p-nitrophenyl nitronyl nitroxide, had the ferromagnetic ordering temperature 0.65 K [17]. More recently, organic molecular soft ferromagnetism was discovered [18] in fullerene $C_{60}$, intercalated with a strong organic donor TDAE, where TDAE stands for tetrakis dimethylamino ethylene $C_2N_4 (CH_3)_8$ . The highest transition temperature for a genuine organic weak ferromagnet is now recorded as 35.5 K [19], but under pressure it is enhanced to 65 K [20]. Mataga's prediction [7] of organic magnetic polymers was realized by a creation of the first plastic magnet consisting of chains of carbon molecules [21].

A comprehensive overview of structural types [22], magnetic properties, new phenomena and opportunities of chemically produced magnets [23]; spin-containing blocks and mechanisms of their interaction [24] is given in an issue of MRS Bulletin "Molecular-based magnets". The advantages of organic and organometallic magnets and their importance for a new millennium are emphasized in a recent review [25]. The mechanism of the intermolecular (through-space) magnetic interactions found in purely organic molecular crystals is reviewed in [Ref. 26]. The paper [27] describes the spin-gap states and the unusual phase transitions observed in novel organic magnetic materials with low-dimensional structures; recent developments in the field of single-molecular magnets can be found in Ref. [28]. Synthesis of organic polymers with intrinsic magnetic properties, namely polyradicals, polycarbenes, polynitrenes is described in the review "Magnetic Polymers" [29].

The difficulties on the way to the production of organic magnets are the follows. The intermolecular exchange tends to align spins of adjacent unpaired s- and p- electrons antiferromagnetically, which is in contrast to d- and f – electrons. Molecular-based magnets consist of assemblies of spin-containing building blocks, but also of ballast non-spin fragments, so the spin concentration is low. Large interspin distances result in small exchange energies and low temperatures of transition to a ferromagnetic state $T_c$. In addition, strong interactions among spins lead to the formation of chemical bonds rather than development of magnetic properties.

Although magnetism has at least a 4000-year history, there is no comprehensive theory that explains the phenomena. One of examples is organic magnetism: along with the phenomena which are understandable and quantitatively described in terms of the models and mechanisms, some aspects need further study and can be even called "unsolved mysteries" [30].

## 3. Room-temperature carbon-based magnets: experimental evidence

In 1996 a paper appeared describing the present state and the future of a new branch of organic magnetic materials – carbon-based magnets [31]. The authors state: "Of many candidates of the magnets, carbon compounds will be the most promising from the practical point of view, because the carbons exhibit a spontaneous magnetization at room temperature and are cheap to make, chemically and physically stable, and easy to process".

Hereinafter we follow mainly the chronological order, describing the carbon based magnetic substances. The papers surveyed in this section meet the following criteria: the substance shows finite magnetization at room temperature, and proof is given for the absence or for negligible amount of metallic impurities. The probability always exists that the claimed high-temperature ferromagnetism of organic substances in actual fact has an extrinsic origin [32]. On the other hand, many facts and especially dependencies set forth below can be explained only by intrinsic properties of the substances.

After the three mechanisms for the stabilization of ferromagnetic coupling in molecular solids have been predicted, several attempts have been made to synthesize magnetic polymers, and interesting magnetic properties have been observed for vinil polymers containing nitroxile radicals in the side chain of the mer unit [33], in galvanoxyl radical [34] and mixture of galvanoxyl and hydrogalvanoxyl; the latter substance is considered as an assembly of segments consisting of galvanoxyl radicals, ferromagnetically coupled to each other [35] The ferromagnetic interaction in galvanoxyl radical was shown to extend one-dimensionally[36 - 39].

## 3.1. Chains of interacting radicals.

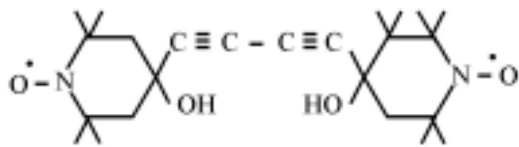

**Fig. 2. Mono-BIPO**

In 1986 the first magnetic polymer was synthesized [40, 41], polydiacetylene crystal poly-BIPO, where BIPO stands for 1,4 bis (2,2,6,6-tetramethyl-4-piperidyl-1-oxyl)-butadiene (Fig.2). This ferromagnetic material is obtained by thermal or photostimulated polymerization of the α-form of mono-BIPO. A spontaneous magnetization typical for ferromagnets was observed, disappearing at temperatures of 420 – 460 K. The value of magnetization obtained in the first experiments was about $M_s$ = 0.02 emu / g, which is only 0.1% of its theoretical value, indicating that only a very small fraction of the material posesses ferromagnetic properties. It is possible to separate magnetic and non-magnetic fractions using the field of a permanent magnet. Ferromagnetic particles with a remanent magnetization above 1 G have thus been obtained. Diffraction patterns reveal the absence of the long-range order and a high packing anisotropy. The ESR spectra detect a strong magnetic anisotropy showing that poly-BIPO is a quasi-one-dimensional organic ferromagnet.

The chemical reaction is complex and the resulting polymer is not very reproducible. Nevertheless, a number or groups succeeded in obtaining ferromagnetic poly-BIPO. Modification of polymerization conditions allowed obtaining a polymer with higher values of spontaneous magnetization [42, 43]. The polymer was subjected to magnetic separation, and particles with magnetizations as high as 0.5 emu/g were obtained. However, no difference in the chemical properties between magneto-attractive and nonattractive parts has been found. Ferromagnetism due to transition metal impurities has been ruled out by the ICP (inductively coupled plasma emission spectroscopy) analysis with a sensitivity of $10^{-5}$ mass %. Obviously, even in the attractive fraction of polymer the percentage of a ferromagnetic phase is very small, and this phase escapes structural detection. Ferromagnetism has also been found in an analogue compound 2,4-bis-(2,2,5,5,-tetramethyl-1-oxyl-3-pyrrolinecarboxylate9-hexadiin [41], poly – BIPENO and also pyro-PAN, indicating a similar origin of magnetism [44]

The experimental evidence of ferromagnetism promoted theoretical design of organic ferromagnets [44]. The main principles have been formulated earlier [45]. The design is based on two rules: all constituent atoms and molecules should be paramagnetic and interaction should be ferromagnetic. In common conditions the intermolecular exchange aligns spins of adjacent s- and p- unpaired electrons in antiparallel (antiferromagnetically). Ferromagnetic exchange interaction can be achieved if the radicals are set in a definite way.

Possible arrangement leading to ferromagnetic interactions is shown in Fig. 3, a. The radicals in the initial chain (white circles) are oriented antiparallelly, and the spins are compensated. If a radical (black circles) is linked chemically to every second radical of the chain, the interaction of "white" and "black" radicals will be also antiferromagnetic. But the whole molecule will have more spins "up" than "down" and hence possess a magnetic moment proportional to the molecule length.

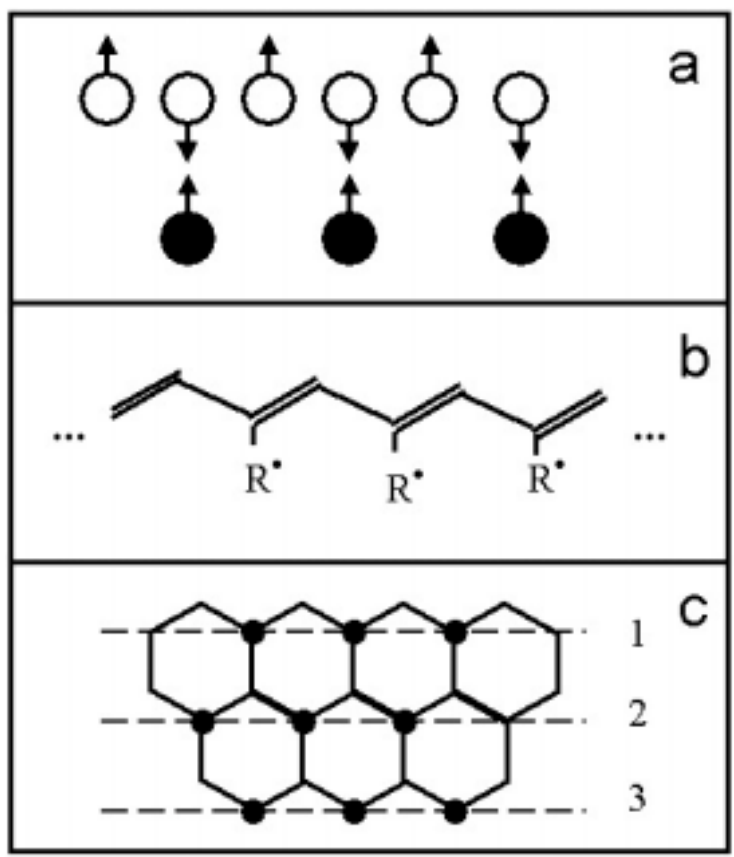

**Figure 3. Theoretical models for organic ferromagnets. a) a chain of interacting radicals; b) an example of one-dimensional polymer system built according to the model; c) a two-dimensional graphite net motif. Adapted from [44].**

Several polymers have been prepared under this prescription, among them ferromagnetic poly- BIPO and poly- BIPENO, described above. The model structure of poly- BIPO is shown in Fig. 3, b.

A very interesting object is a planar graphite-like structure (Fig. 3 c). If some carbon atoms in a graphite two-dimensional network are substituted by three-valence atoms like B, N, Al (black points), the resulting magnetic moment can be extremely high. Similar result is expected for the pure carbon structure, if the black circles represent $sp^3$ hybridized carbon atoms. The atoms in the even lines (1, 3, 5…) should be linked with the atoms in the lower graphene sheet, and the atoms in the odd lines (2, 4, 6…) couple with the upper ones. Thus a stable all-carbon structure with the ferromagnetic interaction can be created. A. Ovchinnikov argues that two $sp^2$ radicals separated by a carbon atom in $sp^3$ state interact ferromagnetically.

A model of a hypothetical three-dimensional carbon structure which meets the above conditions is shown in Fig. 4. The structure has a monoclinic cell of P2/m group with the parameters a = 3.99 Å, b = 5.08 Å, c = 2.59 Å, β = 82.1°. The density of the material is estimated ad 3.07 g / cm$^3$. This three-dimensional magnetic carbon in considered as an intermediate graphite-diamond structure (IGD). The theoretically calculated value for the magnetization of this structure is 230 emu / g.

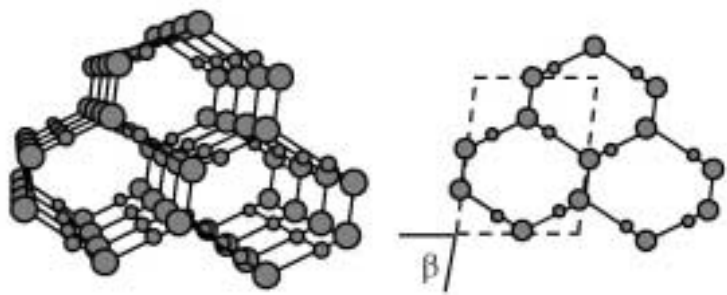

**Fig. 4. An intermediate graphite-diamond structure. Adapted from [44].**

The idea of a three-dimensional ferromagnetic carbon was experimentally realized by A. Ovchinnikov in a PIRO-pan structure prepared by pyrolytic decomposition of polyacrylonitrile (PAN). Different samples showed the value of saturation magnetization from 6 to 15 emu / g. Magnetic separation increases these values to 150 – 300 emu / g, whereas the separated magnetic part distinctly differs from the rest material in being crystalline. This phase is characterized by an extremely high spin concentration ($10^{23}$ cm$^{-3}$) Remanent magnetization is about ten lime less that the saturated value, coercive force lies in the range 80 – 100 Oe and the Curie temperature reaches 500°C. Impurity analysis by means of LAMMA (laser amplified time flight spectroscopy), SIMS (secondary ion mass spectroscopy), electron scanning microscopy, and EES (electron emission spectroscopy) indicated an absence of correlation between magnetic impurities and total magnetization. The main impurities in ferromagnetic carbon are nitrogen and oxygen: 2.3 and 4.7 at. %, respectively, a smaller number of π-electrons than in graphite is in agreement with the theory [46].

An alternative way to produce organic magnets was proposed by J. B. Torrance et al. Reaction of s-triaminobenzene with iodine yields a black, insoluble polymer with ferromagnetic properties [47]. Magnetic moment decreases with increasing temperature indication the Curie point at 700 K. This result is irreversible due to the thermal decomposition of the sample Fig. 5, and this fact itself proves that ferromagnetic signal is intrinsic. The difficulty is that common to all organic magnets: the complex process includes a large number of chemical reactions and is hardly reproducible. The authors suggest a possible structure of the polymer and put forward a model for a ferromagnet containing segregated stacks of radical ions. [48]

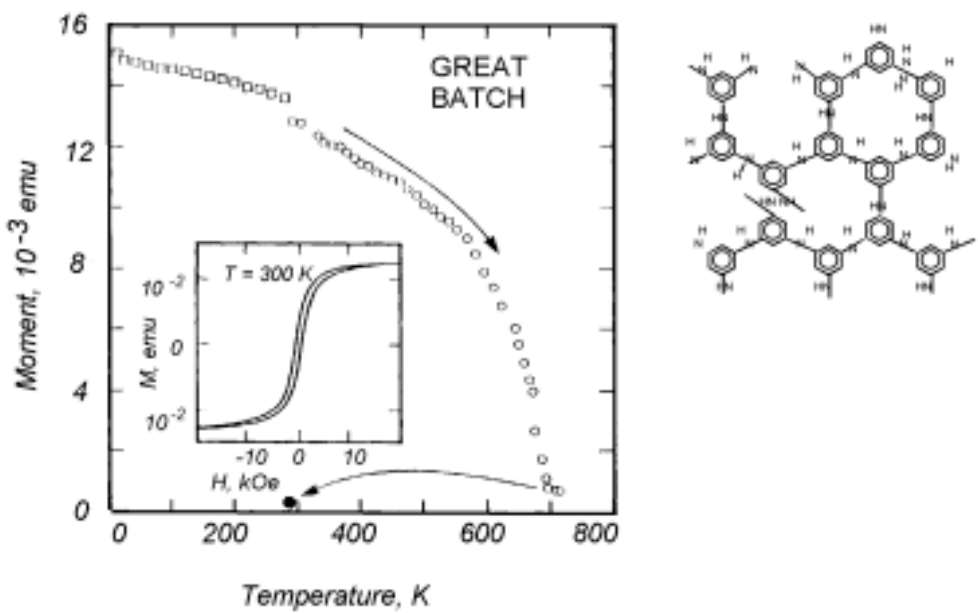

**Fig. 5. Magnetic moment for the best run obtained from acetic acid, measured at 10 kOe using a SQUID (open squares) and a VSM (circles). Note that the room temperature moment disappears by the heating (black point). A possible structure of unoxidized polymer from triaminobenzene is shown in the right [47].**

A different approach for obtaining a ferromagnetic carbon compound was chosen by M. Ota et al. [49, 50]. Whereas the triarylmethane resin is normally a diamagnetic, synthesis of this substance under the magnetic field yields a ferromagnetic substance. This behavior is attributed to the radicals produced by dehydrogenation. The conclusion is supported by the evidence that ferromagnetism increases as the dehydrogenation proceeds.

A synthetic approach modified from the McConnell –Breslow model described in [51] demonstrates that the resulting solid matrix stabilizes the ground high spin state in a wide range of temperatures.

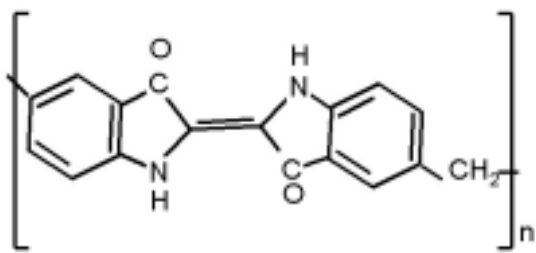

**Fig. 6. Indigo unit in a polymeric chain.**

A sequence of chemical reactions polymerizes indigo units to a chain (Fig. 6), yielding a stable polymer which attracts to a magnet even by heating at 200°C [52]. The room temperature values for saturation magnetization, remanent magnetization and coercive force are $M_s$ = 0.7 emu/g; $M_r$ = 0.08 emu/g; $H_c$ = 120 Oe. Transition metal content was determined by ICP atomic emission spectroscopy to be 26 – 28 ppm Fe and less than 1 for Co and Ni. The $M_s$ value is equivalent to about 3200 ppm Fe, two order of magnitude greater that that determined from ICP. IR spectra show that this material has essentially an indigo skeleton, but the appearance of several additional features suggests that polymer also contains irregular fragments, hydrogen bonds and so on. The main feature of the ESR spectrum is a large broad line due to a high spin state and a sharp line with g = 2.0048 and $\Delta H_{pp}$ = 7 G due to paramagnetic absorption.

## 3. 2. Carbonaceous substances with mixtures of sp$^2$ and sp$^3$ coordinated atoms

Among other organic ferromagnetic materials, pyrolytic carbon is of particular interest. Pyrolytic carbon is highly oriented in structure and the materials pyrolyzed at relatively low temperatures, 600 - 1300°C, contain a large number of unpaired electrons in the graphite skeleton. It is possible to expect spin exchange interactions between unpaired electrons in the graphitic network.

Ferromagnetic pyrolytic carbon with a clear hysteresis loop even at room temperature was obtained by the CVD method using adamantane as a starting material.

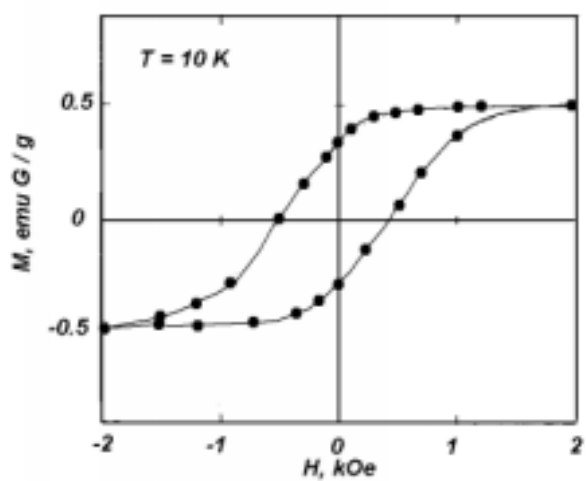

**Fig. 7.  Magnetic field dependence of the magnetization of pyrolytic carbon at 10 K, showing a well-pronounced hysteresis loop [53].**

Adamantane was chosen for obtaining pyrolytic carbon due to its ability to graphitize [53]. The starting material was heated at 1000°C in the Ar atmosphere, evaporated, pyrolyzed and deposited on the inside surface of the quartz tube. This material shows a distinct ferromagnetic loop with $M_s$ = 0.5 emu/g; $M_r$ = 0.35 emu/g; $H_c$ = 600 Oe (measured at 10 K, Fig. 7); these values decrease with temperature, but at 400 K the magnetization still keeps 1/3 of its initial value (Fig. 8).

Assuming that each carbon atom has a spin of an unpaired electron which contributes to the ferromagnetic ordering, the observed saturation magnetic moment is only 0.1% of the theoretical value. The EPR studies reveal rather high spin concentration of $10^{21}$ spin/g at room temperature, and ferromagnetic behavior can be ascribed to the spins of unpaired electrons in the graphitic network. The presence of transition metal impurities was not detected within the 25 ppm sensitivity of the ICP analysis, and ferromagnetism of these samples, produced in a reproducible way, should be considered as an intrinsic property of carbon.

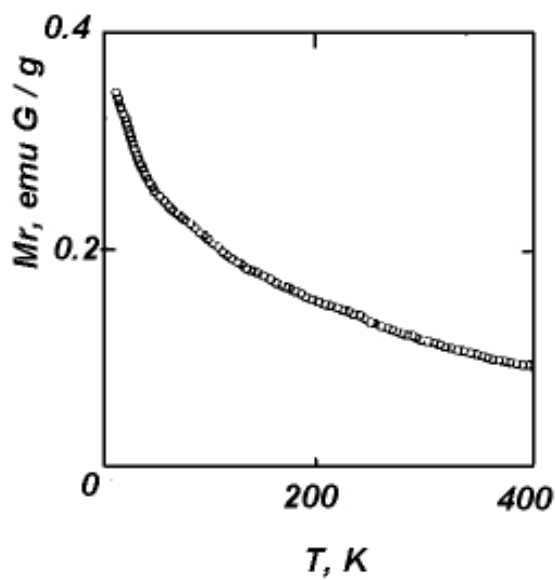

**Fig. 8. Temperature dependence of the residual magnetization, showing ferromagnetism, which is maintained up to 400 K [53].**

Studies of the material structure [54] revealed two conditions which are necessary to obtain a ferromagnetic material: a high concentration of unpaired electrons and a highly oriented structure of the material. X-ray diffraction patterns (Fig. 9) show that ferromagnetic pyrolytic carbon has higher crystallinity than ordinary one. From the ESR measurements, the spin concentration was estimated as $10^{21}$ spin/g and assigned to π-spins due to $sp^2$ radicals.

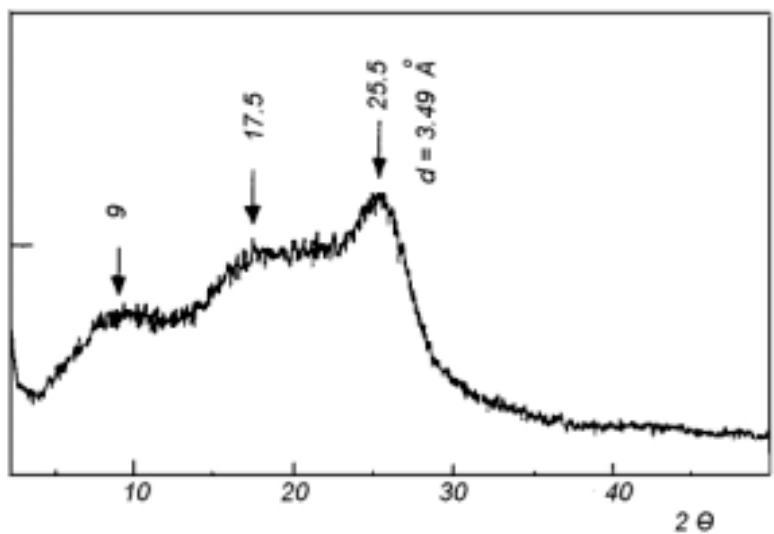

**Fig. 9. X-ray diffraction pattern of pyrolytic carbon [54].**

A thorough examination of the pyrolytic carbon prepared from adamantane was made by another group, in order to check the reproducibility of this process [55]. Two samples were prepared at identical conditions, both were lustrous black in color, insoluble and infusible. The obvious batch dependence was found: both sample showed a sigmoid magnetization curve, but one of the samples was obviously superparamagnetic, while the second one showed the features of ferromagnetic behavior: hysteresis accompanied with characteristic satellites in the EPR spectrum. The saturating behavior in the samples is preserved at room temperature, whereas remanent magnetization disappears at about 200 – 250 K, and the authors conclude that a more finely controlled CVD process will result in the stable room-temperature ferromagnet. The authors [55] emphasize that adamantane-based pyrolitic carbon shows completely different magnetic features compared to other pyrolitic products. They believe that the spins responsible for the peculiar magnetic property come from $sp^3$ type radicals (σ spins) remaining in the product structure. The analysis of ESR spectra suggests a special case of correlated alignments of spins, possibly involving those through the interchain island interaction.

Heat treatment at 1000°C proved to be an effective procedure for obtaining magnetic carbon compounds. These carbon powders were prepared from polymers, such as polyvinyl chloride, phenol resin, and pyrene-benzaldehyde copolymer [56].

Strong magnetic properties were observed for a pyrolytic carbon prepared from cyclodecane by heating at 950°C [57]. The resulting lustrous carbon substance was magnetically separated into two fractions, magnetic A and non-magnetic B. The magnetic fraction proved to be stable in air and preserved its characteristics for at least one month: $M_s$ = 1.07 emu/g; $M_r$ = 0.21 emu/g; $H_c$ = 163 Oe. These values decreased with increasing temperature, but very slow: the values at room temperature accounted about 80% of those at 4 K. The ESR spectra of part A show 2 characteristic signals: a broad line centered at 2940 G and a narrow signal with g = 2.0014 and $\Delta H_{pp}$ = 6.3 G (Fig. 10).

Decreasing temperature results in a decrease in the broad line temperature and in an increase of the narrow line signal with the shift of g-factor to 1.9730. The intensity of paramagnetic narrow line if higher for the fraction B, whereas the intensity of the broad line for the fraction B is much less. The intensities of the broad ESR signal closely correlate with the values of saturation magnetization: the more magnetization, the more pronounced is the EPR broad signal. The shape of the EPR line is characteristic for carbon, but not for metallic impurities. The observed $M_s$ value could be explained by two ways: either 0.2% of carbon atoms possess unpaired spins contributing to a magnetic ordering, or the material contain 4800 ppm of iron. As ICP analysis and fluorescent X-ray technique revealed less than 50 ppm Fe and the absence of Ni and Co, the extrinsic origin of signal can be discarded.
.

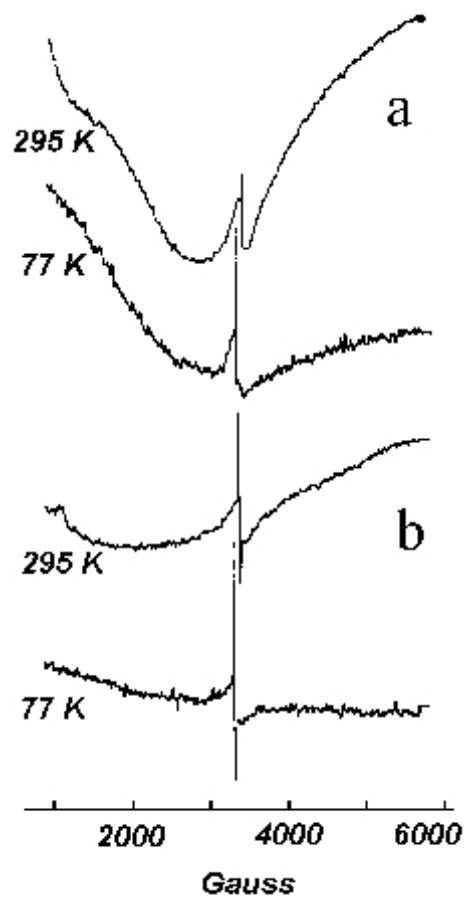

**Fig. 10. ESR spectra of magnetic fraction A (a) and non-magnetic fraction B (b) of pyrolytic carbon. [57]**

### 3.3. Nitrogen-containing carbons.

Magnetic and electrical properties of pyrolyzed polyacrylonitrile (pyro-PAN) has been studied, and possible models of ferromagnetism in this organic substance are presented [58]. A technical PAN ($-CH_2 - CH - CN -)_n$ is a diamagnetic compound, while the product of its pyrolysis is a ferromagnet, slowly degrading with time being exposed to air. The authors consider the formation of ferromagnetic macromolecules with the exchange interaction which belong either to neighboring carbon atoms, or to neighboring nitrogen atoms. The deterioration of magnetic properties is considered as a result of paring of electron spins, participating in the magnetic ordering, with the free spins of oxygen molecules

Magnetization curves exhibited hysteresis loops at room temperature for the product obtained by the pyrolysis of the mixture of phenyldiamine and melamine: hydrocarbons that also contain nitrogen atoms in its molecular structure [59]. Pyrolysis was made at different temperatures, and the product pyrolyzed at 600°C gave a maximum spin concentration of $10^{19}$ spin / g and the highest values of saturation magnetization, remanence and coercive force, $M_s$ = 0.624 emu G / g, $M_r$ = 6.65 $10^{-2}$ emu G / g, $H_c$ = 125 Oe. The molecular structure of the obtained polymeric compounds is complex, and the results cannot be interpreted in a simple way. The existence a remarkable isomer effect of PDA on the magnetic properties prompts that the positions of radical substituents are important for the appearance of parallel spin orientation.

The authors made a comparative study of several organic compounds. Being pyrolized at 600 C, the products obtained from 5 different compounds were attracted to a magnet and exhibited hysteresis loops at room temperature. The best effect was obtained for guanine (a base of DNA in cellular tissue) pyrolyzed at 600°C [60]. The saturation magnetization in this compound is $M_s$ = 0.624 emu G / g, and the spin concentration is estimated as $4.25 \cdot 10^{19}$ spin/g. the origin of magnetism is interpreted in terms of interacting nitrogen radical spins. The coexistence of ferromagnetic and paramagnetic spins suggests that ferromagnetic clusters are dispersed in a sea of paramagnetic spins.

A mixture of 2,4,6-triphenoxy-1.23,5-triazine (TPTA) and melamine (M) pyrolized at 600°C was shown to be ferromagnetic with a room-temperature saturation of $M_s$ = 0.3 emu G / g [60]. Treatment at 950°C in Ar atmosphere including a trace of dry air leads to a creation of a very strong organic ferromagnet [61].

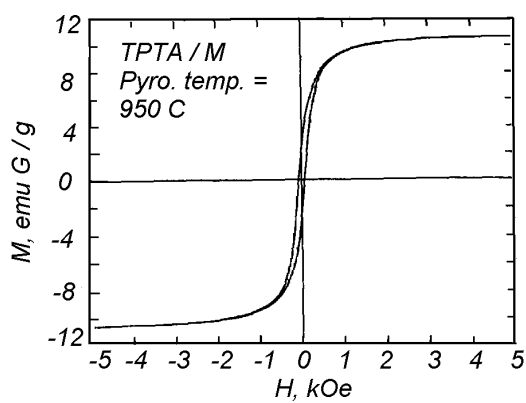

Fig. 11. The magnetization curve at room temperature of a darkred material prepared by pyrolysis of a mixture of TPTA and M at 950 C [61].

Fig. 11 shows the M – H curve of the darkred sample, prepared by pyrolysis at 950°C in Ar atmosphere including a trace of dry air. This curve demonstrates extremely high saturation magnetization $M_s$ = 10.4 emu G / g, as well as remanent magnetization $M_r$ = 2.6 emu G / g and coercive force 72.2 Oe. Fig. 12, a shows ESR lines of this material measured at the temperatures of 100 and 290 K. The lines are characterized with a very high intensity, corresponding to the spin concentration of $4.7 \cdot 10^{23}$ spin / g, and extremely wide peak-to-peak linewidth, $\Delta H_{pp}$ of 804 G at 290 K and of 1570 G at 100 K. The peak height decreases with decreasing temperature, but the integrated line intensity remains constant due to the broadening of the linewidth at low temperatures.

Temperature dependence of magnetic susceptibility $\chi$, estimated from the ESR line, did not show paramagnetic-like behavior in the temperature range 100 – 300 K, suggesting that the Curie temperature is higher that 300 K. The dependencies of g-factor and $\Delta H_{pp}$ are shown in Fig.12, b. Both values decrease as temperature increases, whereas for three-dimensional ferromagnet g-factor should be independent of temperature, and $\Delta H_{pp}$ should decrease with decreasing temperature. The magnetic behavior of this organic compound is thus inconsistent with the behaviour of convenient ferromagnets.

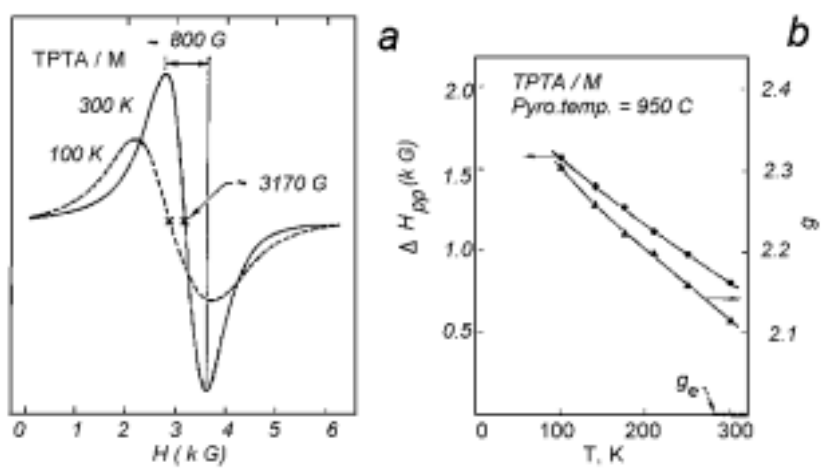

Fig. 12. a: ESR line of the darkred material prepared by the pyrolysis of the mixture of TPTA and M at 950 C in Ar atmosphere including a trace of dry air (or oxygen). Spectra are taken at 100 and 300 K.  b: Temperature dependence of the peak-to-peak linewidth $\Delta H_{pp}$ and the g-factor, determined from the ESR line of the darkred material. Adapted from [61]

### 3.4. Carbons from hydrogen-rich starting material

The values of saturation magnetization for amorphous-like carbons prepared from tetraaza compounds by the chemical vapor deposition (CVD) method [62] are among the highest reported.

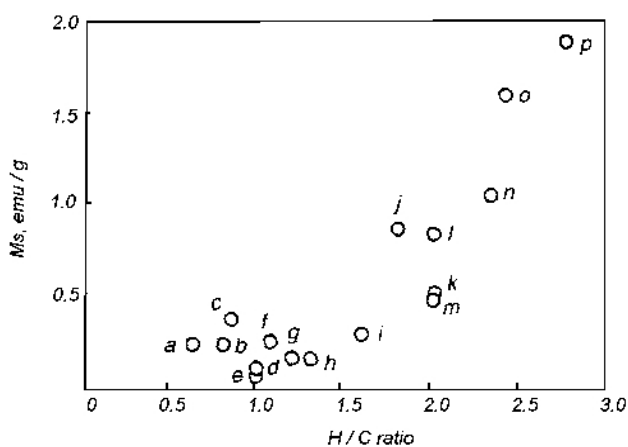

Fig. 13. Dependence of saturation magnetization on the hydrogen-carbon ratio in the starting material. Letters stand for different organic compounds, see text. [62]

Aza-carbon is the product of pyrolysis of commercially available tetraaza compounds at 950°C. According to the X-ray diffraction, electron diffraction analysis and Raman spectroscopy, aza-carbon does not show distinct crystalline phase. The magnetic and ESR properties are notable for its reproducibility. The $M_s$ value reaches 2.56 emu / g, or 0.005 $\mu_B$ / carbon atom, pointing out that only 0.5% of carbon atoms contribute to the ferromagnetic ordering, The ESR spectra show two characteristic lines: a broad line over the range 1000 – 6000 G and a sharp signal with g = 2.0013, $\Delta H_{pp}$ = 6.1 G.

K. Murata et al. [62] analyzed the relationship between the hydrogen content in the starting material and the value of saturation magnetization at room temperature. They studied the following substances: *a:* pyrene; *b:* trans-1.2-di(thienyl)ethylene, *c:* triphgenylmethane; *d:* paracyclopane; *e:* phenol; *f:* tetrahydrocarbazole; *g:* dibenzo-18-crown-6; *h:* octahydroacridine; *i:* adamantane; *j:* tricyclohexylmethanol; *k:* cyclododecanol; *l:* cyclodecane; *m*: cyclopentadecane; *n:* n-hexane; *o:* 1,4,8,11-tetraazacyclotetradecane; *p:* 1,5,8,12-tetraazadodecane. The results are shown in Fig. 13 as the $M_s$ value versus hydrogen-to-carbon ratio in the starting material. It is clearly seen from the picture that high hydrogen content is favourable for the magnetic properties, and this is especially clear when the H / C ratio exceeds 2. The authors speculate that the atomic hydrogen generated from the aza-compounds play an important role in formation amorphous-like carbons rather that graphitic.

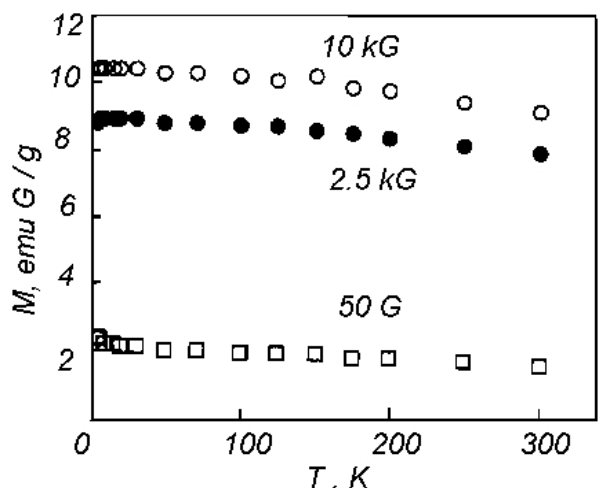

**Fig. 14. Temperature dependence of magnetization of the carbonaceous material at 10.0, 2.5, and 0.05 kG. [63]**

Having found that the $M_s$ values of the ferromagnetic properties depend on the hydrogen content in the starting material, Murata et al. [63] extended their research and found that 1,2-diaminopropane having a H / C ratio of 3.3 is an excellent precursor for a strong carbon magnet. A strong and air-stable carbon magnet is obtained by the rapid pyrolysis at 950 C. Analysis by powder X-ray diffraction, electron diffraction, and transmission electron microscopy have shown that the material is amorphous. The magnetization value is 10.5 emu G / g, corresponding to 0.022 $\mu_B$ / carbon atom. The temperature dependence of magnetization shows non-Curie-Weiss behavior: it is very slightly decreases with temperature. The Curie temperature is estimated as $T_c > 500$ K, but the exact value is not determined due to decomposition of the structure at about this temperature.

Some structural information of the material can be obtained from the FTIR ATR measurements, which prove that the material comprises a mixture of the $sp^2$ and $sp^3$ carbons.

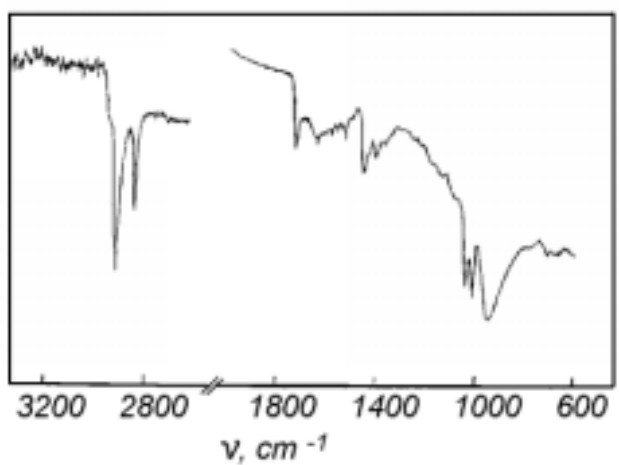

**Fig. 15. The FTIR ATR absorption of the carbonaceous film in the 600 to 3200 cm$^{-1}$ region [63].**

The FTIR ATP spectra of carbonaceous film, deposited on a quartz substrate, exhibit two bands in the 2850 – 3000 cm$^{-1}$ region, and a number of bands at 850 – 1750 cm$^{-1}$, associated with the combined absorptions of the C – H stretches and C – H in-plane and out-of-plane bends of $sp^3$ (-CH2-) and $sp^2$ (=CH-) carbons. Other evidence of the mixed $sp^2 / sp^3$ structure is: (1) the presence of the aromatic nitrile and heterocyclic compounds, indicating that aromatization and /or heterocyclization occurs in the process; (2) the values of electrical conductivity are much lower than for graphite or $sp^2$ carbon; (3) STM and AFM observations demonstrate an irregular array of carbon atoms of benzenoid rings. Evidently, the origin of high magnetization values and high stability of the material is due to a three-dimensional network structure consisting of both $sp^2$ and $sp^3$ carbons. The formation of the $sp^3$ carbons is accelerated by the atomic hydrogen generated in situ from the starting material (diamine) under the chosen experimental conditions.

The presence of a permanent magnetization in the described above amorphous-like carbons prepared from nitrogen-containing starting material is controlled by two major factors: starting material and technological process. Wide range of starting material was investigated by H. Ushijima et al. [64], and the properties of ferromagnetic or superparamagnetic carbons have been described. These carbons do not have distinct crystalline phase, as shown by powder X-ray diffraction, Raman spectroscopy, transmission electron microscopy, and electron diffraction analysis. X-ray photoemission spectroscopy (XPS) reveals minor differences in the $C^{1s}$ binding energy and full width at half maximum between the amine-derived carbons and commercially available graphite, which are consistent with the presence of a disordered structure.

Clear evidence that the amorphous structure is closely related to the magnetic properties, comes from the XPS spectra of plasma-treated and untreated carbon surfaces. The C1s energy loss region is subdivided into 3 components: amorphous-like carbon, graphite and diamond. The ratio of these components was measured for the plasma treated samples at different power, and this ratio was compared to the magnetization values. Fig. 16 shows clear the one-to-one correspondence between the data.

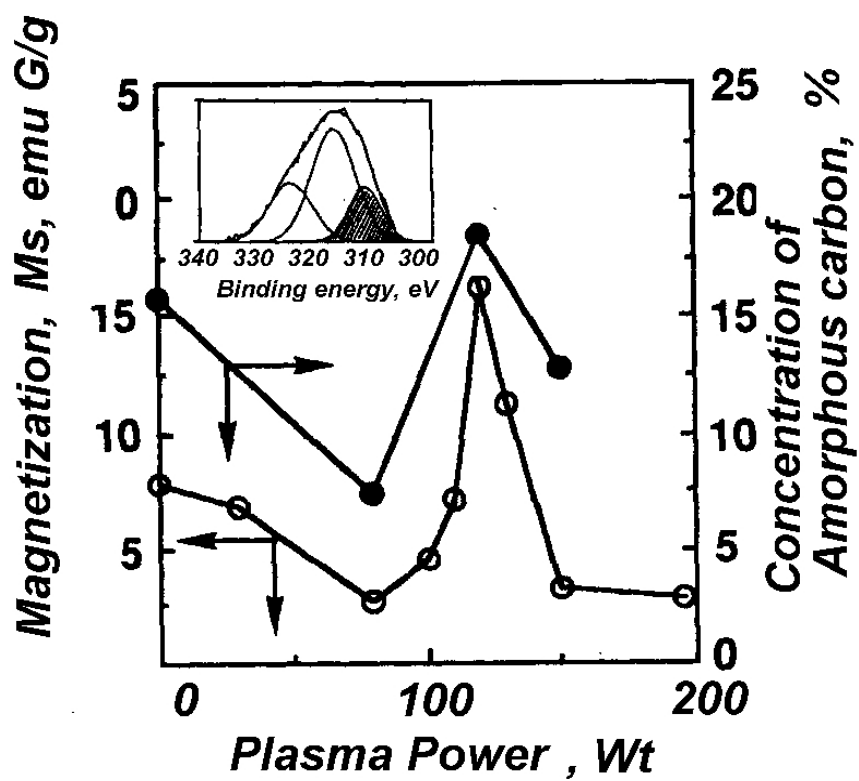

Fig. 16. Plasma power dependence of (1 – solid circles) the $M_s$ value at room temperature and (2 – open circles) amorphous carbon content of the carbon material. Inset: an example of analysis of $C^{1s}$ energy loss spectra. Adapted from [64].

A different model for the carbon magnetism is proposed by R. Setnescu et al. Studying the ferromagnetic phase of pyrolysed polyacrylonitrile, they suggested that magnetic properties are related to the existence of chemical species such as nitroxide radicals, and radicals grafted on the carbon atoms belonging to graphite-like structures [65].

There have been several attempts to repeat the results of K. Murata et al. It has been shown that the nature of deposited carbonitrides, their elemental composition as well as their structural organization mainly depend on 2 experimental parameters: (1) the way to afford the excess of energy for getting a metastable state and (2) the nature of the precursors. The number of unpaired spins is shown to be a function of the chemical reactivity under plasma constraints [66]. M. Trinquecoste et al. studied magnetic properties of carbonitride films elaborated by plasma enhanced chemical vapor deposition. The phase obtained in this work, is not of graphitic type but a crystalline $C_3N_4$, and for this material no long range magnetic order is detected. The authors argue that a bulk magnetic ground state is quite unlikely in these compounds, and support this statement with the calculation of a ferromagnetic exchange interaction *J*. In a graphitic system with $10^{-2}$ to $10^{-3}$ spins per carbon atom and a mean distance between spins of about 10 Å a magnetic transition temperature would occur at a temperature lower than 1 K. The authors, however, suppose that there is a possibility to observe an intrinsic magnetic behavior in the systems with non-homogeneous distribution of unpaired spins (as in spin glasses) and with a high surface concentration. The surface effects can be due to magnetic centers, associated with dangling bonds.

### 3. 5. Ferromagnetism of conducting polymers

Conducting polymers consisting of long carbon-based chains composed of simple repeating units have attracted the attention since the early 1980s. They are different from other polymeric materials (which become conducting by the incorporation of carbon black or metal particles) in having an inherent electrical conductivity. The Nobel Prize in Chemistry 2000 was awarded for the discovery and development of conductive polymers.

Conducting polymers such as polyaniline (PANI) have been attracting considerable attention due to the high conductivity achievable by doping, and also due to the ability to yield a ferromagnetic spin alignment upon doping [67, 68]. Ferromagnetic interaction was shown to exist in poly-(*m*-aniline) on the basis of the temperature dependence of the ESR signal intensity, the observation of the forbidden $\Delta m_s = \pm 2$ signals, saturation magnetization and the Brillouin function fitting [69]. To achieve extensive spin stabilization, the synthesis of polyaniline was made with the use of a large size functionalised dopant namely 2-6 quinone disulphonic acid. The obtained conducting emeraldine salt gets attracted to a permanent magnet at room temperature [70]. The growth conditions were shown to be determinative for the magnetic ordering: a competition between ferromagnetic (layered growth) and antiferromagnetic (random nucleation) yields a material with spin ordering.

A ferromagnetic behavior at room temperature is reported in metal-free conducting samples of $ClO_4^-$ doped poly (3-methylthiophene) [71, 72]. Determination of magnetic impurities was performed using a graphite furnace atomic absorption spectrometer (Varian AA 80). The following contamination levels were determined: 1 ppb of Fe, 2 ppb of Ni, and absence of Co. The magnetic properties of the polymer are strongly dependent upon preparation conditions. An interesting effect of changing the coercivity with the water content in the solvent is observed, whereas remanence and saturation remain unchanged. The samples are unstable when stored in air: in five months the ferromagnetic behavior disappears leaving only a diamagnetic contribution. The authors associate the magnetic moment with spin ½ positive polarons, which interact through a Dzyaloshinski – Moriya anisotropic superexchange via the dopant atoms.

### 3. 6. Carbon magnets: role of impurities

The most important question in the study of magnetic properties of carbon is whether its nature is intrinsic or extrinsic. The papers described in Section 3 report that the iron concentration in the samples is between 10 and 20 ppm, whilst the cobalt and nickel concentration is less. Being homogeneously distributed in the carbon matrix, the iron impurities cannot contribute to the magnetic ordering due to the large interatomic distances.

One should, however, consider the possibility of clusterization of metallic atoms. If we assume that all iron impurities in the samples are concentrated in a cluster with $Fe_3O_4$ composition (the presence of metallic iron is unfeasible), the contribution of such a cluster to the magnetization values is estimated as 0.02 emu / g.

In the majority of the papers the magnetization values substantially exceed this value, in some papers the magnitudes are comparable. An important moment is that there is no difference in the impurity content among the samples examined, but the magnetic properties are strongly dependent on the preparation conditions: reaction time, temperature and the atmosphere. There is also substantial dependence on the nature of starting material: high hydrogen content is desirable, and the branched isomers are favourable.

For example, organic ferromagnetism of the pyrolized polyacrylonitrile was studied with the emphasis on the careful chemical analysis of the samples [73]. Among ten measured samples, one showed the saturation magnetization at least seven times bigger than it would be produced by the iron impurity, whereas the sample with ten times more iron concentration showed 50 times less magnetization. Fig. 17, plotted using the data from Ref. [73], clearly shows an absence of correlation between the magnetic properties and iron impurities in the sample.

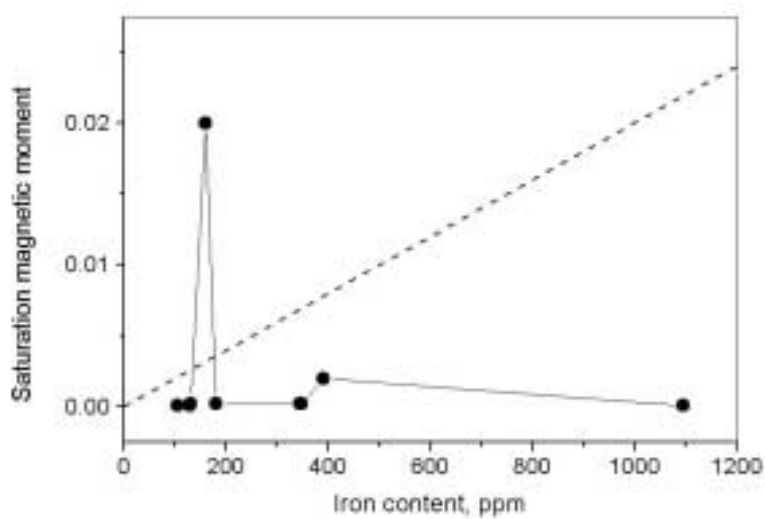

**Fig. 17. Saturation magnetization of the pyro-PAN samples versus iron impurity content. Points: experiment. Dashed line: expected saturation magnetization, calculated under the assumption that all iron impurities contribute to the ferromagnetic ordering. Plotted using the data of [73]**

In another study, magnetic properties of Fe / activated carbon powder were examined [74]. If the iron concentration is less than 6%, the magnetism of Fe / activated carbon is identical to that of pure activated carbon. Iron-induced ferromagnetism could be observed when the Fe concentration exceeds 6-8%.

It is highly desirable to produce totally metallic-free organic substances. However, contamination of iron is often unavoidable, and it is significant to know the effect of tiny amounts of metal to the magnetic properties. The effect of trace iron on the magnetic properties of the fine particles prepared by laser-induced ablative decomposition of solid perylenetetracarbohylitc dianhydride (PDA) was thoroughly studied in Ref. [75]. Quantitative analysis of Fe, Co, Ni, Cr and Mn in PDA and reaction products (PRO) was made by X-ray fluorometry (XRF) and by inductive coupled plasma atomic emission spectrometry (ICP-AES). The content of Co, Ni, Cr and Mn was of the order of 1 ppm. The iron concentration varied from 3.5 to 2200 ppm. Sudden change in the magnetic behaviour of PRO, from diamagnetic to paramagnetic, was observed between the iron content of 2000 and 2100 ppm.

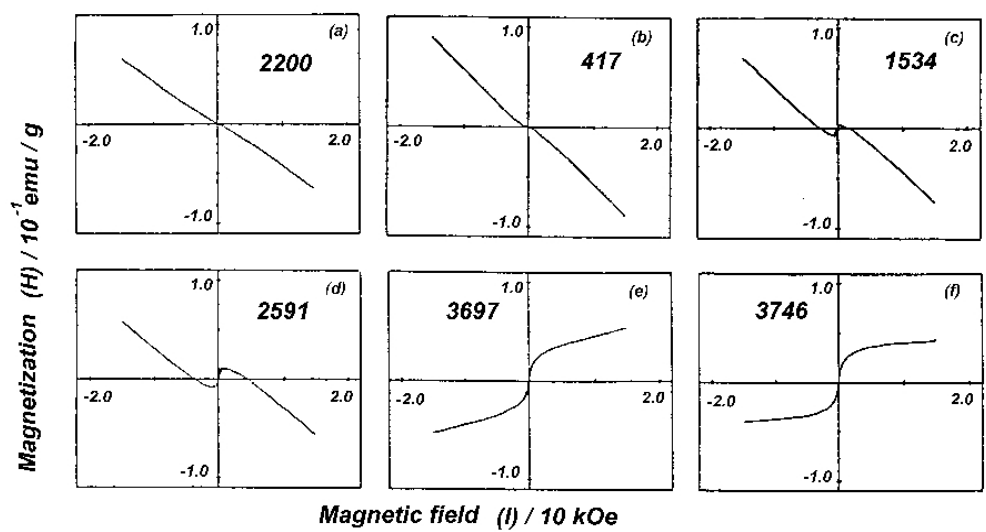

Fig. 18. Vibrating sample magnetometer spectra of PDAs and PROs with different iron content. (a): The spectra of PDA 3.5 – PDA 2200 are identical; (b) PRO 295; (c) PRO 900; (d) PRO 2000; (e) PRO 2100; (f) PRO 2220. Numbers near the name of the sample: the amount of iron in the starting material. Numbers inside the plots: actual amount of iron in the sample (ppm). Adapted from [75].

It is important to notice that the changes in the magnetic behavior are observed only in the reaction products of the PDA (Fig. 17, b – f). PDA itself remains diamagnetic for the iron content of 3.5 – 2200 ppm (Fig. 17, a), and also shows no absorption in ESR. In contrast, broad absorption lines over the range 100 – 600 mT are observed in the iron-containing reaction products.

### 3.7. Carbon magnets: industrial applications.

The origin of magnetism in carbon-based magnets has not been yet established: the question of where the spins come from and how they couple does not have an unambiguous answer. It seems, however, that the engineering problem of designing a molecular magnet in a predetermined way is much easier than the scientific problem of predicting the structure of a ferromagnetic molecular material.

Magnetic carbon materials are now produced on an industrial scale and are suitable as toner for copiers and other business machines, or as magnetic ink. A brief scan of some of the patents, which describe manufacturing strong carbon magnets free from transition metals, can shed light on the structure / property relationships.

The common feature of different approaches is heating various organic compounds by steps for obtaining paramagnetic amorphous carbon with subsequent graphitization.

K. Murata describes 40 examples of preparation organic materials absolutely free from a ferromagnetic metal [76]. The starting materials have a specific structure and belong to four groups: (a) an aromatic polymer in which the aromatic compounds are connected via alkylene chains, which can be substituted by an aryl, alkyl or aralkyl group; (b) a triarylmethane polymer; (c) a polymer of an acetylene compound; (d) a polymer containing 1-10 hydrogen atoms per 100 carbon atoms. These substances are subjected to a thermal treatment in vacuum or in an inert gas atmosphere. In the absence of some oxidative gas the polymer is decomposed and converted into a carbonaceous substance in the mid-course of graphitization. In the X-ray diffraction analysis a diffraction line (002) showing absorption of carbon is detected in the vicinity of $2\Theta = 25°$. The lattice spacing corresponds to 3.5 – 3.6 Å, showing that the spacing is sufficiently broader than 3.354 Å of graphite. In the next inventions [77, 78] Murata increases the H : C ratio to 1.7 : 1 and uses aliphatic hydrocarbons, alicyclic hydrocarbons, or azo compounds. With the same aim of creating the substance on mid-course to graphitization, he obtains 12 compounds characterized by the coercive force 90 – 120 Oe, and saturation magnetization reaching 9.37 emu /g.

H. Ueda [79] suggests the idea of forming a conjugated double bond by depolymerization of an organic polymer. Polyvinilchloride is heated by steps under the nitrogen flow, dehydrogenated, and quenched to form paramagnetic C. Carbonization of this material at 520°C in vacuum yields an ungraphitized carbide with ferromagnetic properties: $M_s = 3$ emu/g, $M_r = 0.6$ emu/g, $H_c = 400$ Oe. He also treats organic substances (e.g., PVC) at 250 – 800°C in the presence of a halogenating agent and obtains carbonaceous ungraphitized carbide, which proves to be a metal-free strong magnetic material [80]. The presence of a halogen-containing material, such as carbon tetrachloride, helps eliminating hydrogen atoms during the heat treatment. Ueda develops further the idea of obtaining carbonaceous ferromagnetic materials, consisting substantially from carbon alone [81]. He carbonizes specific organic compounds, like phenasine and indigo, to eliminate hydrogen atoms from the starting aromatic compound as completely as possible. Graphitization of the carbonization product does not take place. The resulting material is an almost pure carbon, and its characteristic feature is that σ-unpaired electrons are left therein in an unbonded state. The subject of the discovery is a process which allows one to eliminate hydrogen from aromatic ring compound, but not causes the graphitization of the carbonized material. The residual hydrogen atoms substantially diminish ferromagnetic properties.

Ferromagnetic organic polymer compounds can be prepared from nitrogen-containing starting materials: a highly molecular polymer with a molecular weight of 500 – 500 000 is obtained by a thermal treatment of a cinnamic nitrile derivative polymer [82]. Polymerization of aminoaromatic compounds yields copolymers containing aminoaromatic groups having magnetic properties with the Curie temperature of 350 K [83].

S. Nakajima [84] creates an organic polymer, which has a molecular weight of at least 1000 and is characterized by a high density of verdazyl radicals (over $10^{17}$ g$^{-1}$). Verdazyls is a family of stable delocalized radicals, which is now widely used as building blocks for magnetic materials. The polymer is water-soluble and has a property of responding to a magnet. This substance is used in magnetic ink in the ink-jet printers, thermal-transfer ribbons, writing implements. The image obtained by using the magnetic ink can be magnetically read, thus giving a possibility to use it for printing money, certificates and secure documents. Another polymer, which consists of repeated units containing oxygen, sulfur and NH groups, is applied as a magnetic toner for development of the image [85]. An organic material for magnetic tape, photomagnetic memory of ink is also produced by photooxidizing an organic substance triarylmethil [86].

Y. Ushijima [87] synthesized an organomagnetic material, which consists of amorphous carbide on the way to graphitization, and contains phosphorus and / or boron, and also hydrogen. The saturation magnetization of this carbon substance is 59.8 emu G / g, i.e., equivalent to those of a metal nickel.

Ferromagnetic material can be also obtained from fullerenes. Dispersion of $C_{60}$, $C_{70}$, $C_{76}$, $C_{84}$ fullerenes in an organic polymer or in a non-conducting liquid forms a light processable magnetic material [88]. In another process, fullerenes are doped by halogens from alkali halides, for example, LiF, and ultrasonically dispersed in an organic polymer [89].

Summarizing the survey of the patents, we can conclude that nearly all the materials described meet the requirements of the theory of A. Ovchinnikov [9, 44]. This theory predicts that a ferromagnetic ordering of spins could exist in π delocalized systems due to a lone electron pair on a three-valence element: nitrogen, phosphorus, aluminum, boron, etc. Another possibility of such an ordering is a pure carbon material having an intermediate structure between graphite and diamond, where the alternating $sp^2$ and $sp^3$ carbon atoms play the role of different valence elements. The organic ferromagnetic materials described in this section fall into two categories: the first one is the carbonitride phases or boron / phosphorus carbon compounds, and the second is the pure carbon material on the mid-course to graphitization.

## 4. Magnetic properties of graphite

### 4.1. Bulk graphite.

The diamagnetism of the $C^{4+}$ ion is $-1.2 \cdot 10^{-8}$ emu/g [90]. Bulk graphite has a c-axis diamagnetic susceptibility of $\chi_\perp = -(22 - 50) * 10^{-6}$ emu / g which is caused by the itinerant π-electrons in a semimetallic structure. Magnetic properties are different for the hexagonal and rhombohedral forms [91], rhombohedral form being more anisotropic. The temperature dependence of magnetic susceptibility of graphite was calculated by J. W. McClure [92, 93] taking into account strong interband effects. The orbital diamagnetism of disordered graphitic materials is described by A. S. Kotosonov [94]: The large value of diamagnetic susceptibility is attributed to the broadening of the energy spectrum due to the randomness. It is usually accepted that when the magnetic field is applied parallel to the basal plane, the graphite susceptibility is also diamagnetic: $\chi_{//} = -5 \cdot 10^{-7}$ emu/g. However, there is experimental evidence that this component is paramagnetic $\chi_{//} = 2 \cdot 10^{-6}$ emu/g [95]. Negative experimental values of the in-plane susceptibility can be due to sample misalignment, since the c-axis component is strongly diamagnetic. Thorough experiments show that the low-field dependence of the moment on the field is non-linear not only for graphite and fullerene $C_{60}$ [95, 96], but also for diamond, and carbon nanotubes [96].

There exist both theoretical predictions and experimental evidence that electronic instabilities in pure graphite can lead to ferromagnetic and superconducting properties even at room temperature. Magnetoresistance measurements on highly oriented pyrolitic graphite (HOPG) suggested the occurrence of superconducting correlations below 50 K [97], whereas graphite-sulfur composites demonstrate a clear superconducting behavior below the critical temperature $T_c$ = 35 K [98] or 37 K [99], and coexistence of superconductivity and ferromagnetism in the graphite-sulfur system [100].

Y. Kopelevich et al. identified both ferromagnetic and superconducting-like magnetization hysteresis loops in HOPG samples below and above room temperature [101]. This is the first experimental work, clearly showing high-temperature ferromagnetic-like behavior and possible "hot" superconductivity in graphite. When the field is applied parallel to basal planes, the M(H) hysteresis loops are ferromagnetic-like, and m(H) = M(H)·V is proportional to the sample volume V, demonstrating that the ferromagnetic behavior is a bulk property of the sample. Heat treatment at 800 K in the He gas atmosphere strongly enhances the magnetic signal (Fig. 19). For fields applied parallel to the basal planes, the main contribution is diamagnetic, but subtraction of a linear diamagnetic background reveals superconducting-like hysteresis loops.

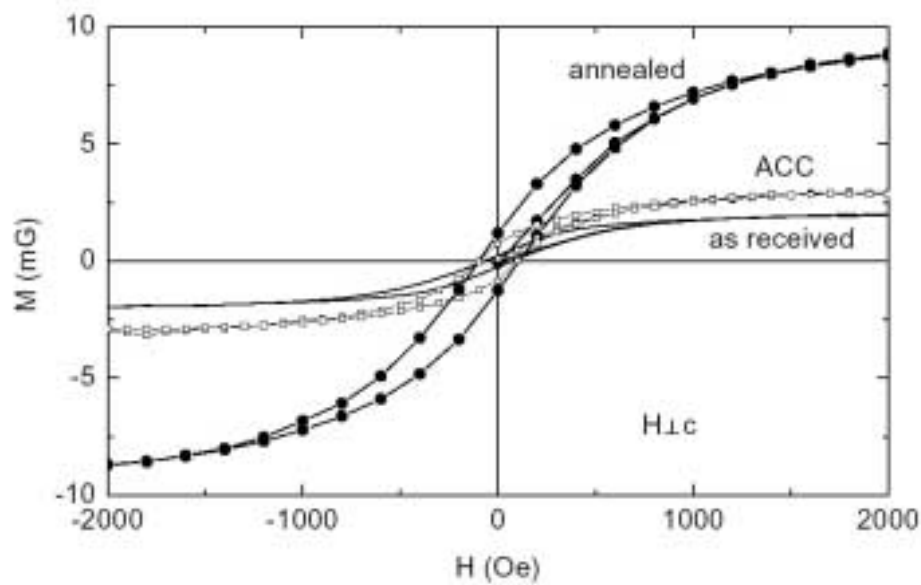

**Fig. 19. Ferromagnetic-like M(H ⊥ c) hysteresis loop measured for:**
- HOPG before annealing (continuous line, T = 300 K)
- HOPG after annealing at 800 K for 2 hours, (black circles, T = 350 K);
- as-received sample from Advanced Ceramic Corporation (ACC, open squares,
T = 300 K) [101].

Recent studies of the properties of bulk graphite by means of conduction electron spin-resonance using S, X and Q microwave bands give a cogent argument of the existence of an effective internal ferromagnetic-like field in graphite [102]. Weak ferromagnetism in various, well characterized samples of HOPG, Kish graphite and natural graphite was studied by P. Esquinazi et al., [103]. An absence of correlation between saturation and remanent magnetization, coercive force and the iron concentration gives a conclusive proof for the intrinsic nature of the ferromagnetic signal. AFM studies suggest that topological defects can contribute to the peculiarities of the graphite electronic properties and thus give rise to ferromagnetic correlations. Also, the magnetic properties can be explained by itinerant ferromagnetism of dilute two-dimensional electron gas with the strong Coulomb interaction between electrons.

## 4. 2. Graphite edges

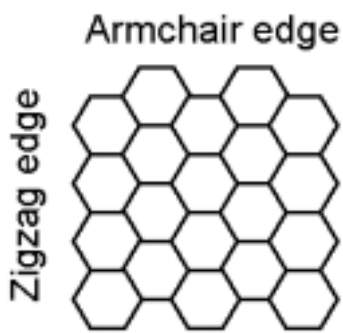

Fig. 20. Types of graphite edges.

Electronic properties of finite graphene sheets are drastically different from that of bulk graphite. If graphite has a stepped surface, a localized state at the Fermi level appears, which is caused by the cut of the graphene layer and is localized near the step. [104]. For the nanosized graphite, edge states are shown to play a leading part [105]. In general, the origin of the spins in carbon materials was attributed to the dangling σ bonds. A different origin for the spins in carbon materials has lead to the detection of novel molecular magnetism in combination with conduction electrons, and predictions of a possibility of ferromagnetism.

The edge of a graphite sheet can be described as a combination of two types of edge shapes: Zigzag and armchair shapes for a graphene sheet are shown in Fig.20. Graphite ribbons are characterized by the edge type and also by the width N: number of lines between two edges. The armchair ribbons are metallic for N = 3m − 1 (m is integer), whereas the zigzag ribbons are metallic for all N. A remarkable feature of the zigzag ribbons is the appearance of a sharp peak in the density of states at the Fermi level [106]. These states do not originate from the bands of infinite graphite, but appear as localized states around the Fermi level where the π- and π*- bands touch in 2D graphite. The examination of the charge density distribution has shown that the electronic state in the almost flat bands is characterized as strongly localized states near zigzag edges. The dependence of the diamagnetic susceptibility on the edge shapes is found from the analysis of the ring currents, which are very sensitive to the lattice topology near the edge [107]. The edge states are present also in the graphite ribbons with arbitrary edges, i.e. with a combination of zigzag and armchair edges: several zigzag sites enable the appearance of an edge state.[108]

As the edge states produce a peak at the Fermi level, they contribute to the Pauli paramagnetic susceptibility which competes with the orbital diamagnetism: for nanographite ribbons with zigzag edges a low-temperature Curie-like behavior changes to a diamagnetic behavior at higher temperatures [109].

Fig 21 [110] shows the temperature dependence of the total susceptibility $\chi = \chi_{orb} + \chi_p$. The total magnetic susceptibility $\chi$ is the sum of the orbital diamagnetic response $\chi_{orb}$ due to the itinerant nature of the n electrons and the Pauli paramagnetism due to their spin. The inset of Fig. 21 shows the contribution of the Pauli susceptibility for the various values of ribbon width N. At low temperatures $\chi_p$ has the Curie-like form, which is originated from the DOS of the edge states.

Similar behavior of magnetic susceptibility is found experimentally on the graphitized nanodiamond [111] and on the pitch-based activated carbon fibers (ACF) with the dangling bond spins associated with 10 - angstrom micro-pores. [112].

The zigzag edge was shown to have peculiar magnetic properties. The presence of almost flat bands leads to the electron-electron interactions which in turn lead to magnetic polarization and to a lattice distortion due to the electron-phonon interactions. Examination of the electron-electron interaction effects in the Hubbard model reveals a possibility of spontaneous magnetic ordering in nanometer-scale fragments of graphite. Huge magnetic moments are induced at both edges by weak Hubbard U and are coupled antiferromagnetically with each other. The existence of local ferrimagnetic structure is possible for a zigzag ribbon, and the emergence of a magnetic order is dictated by the competition with the lattice distortion induced by an electron-phonon interaction. [113].

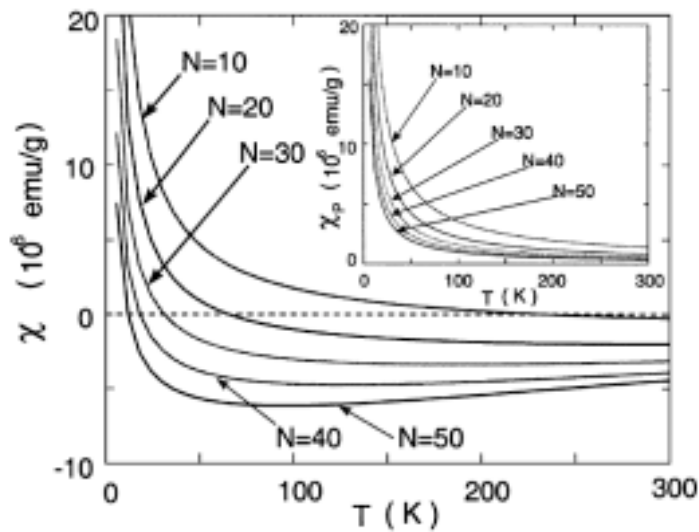

**Fig. 21. Temperature dependence of magnetic susceptibility for the zigzag graphene ribbons [113]**

### 4.3. Magnetism of ACF

Activated carbon fibers are microporous carbons with huge specific surface areas. ACFs can be considered as a three-dimensional random network of nano-graphitic domains with characteristic dimensions of several nanometres. There exists strong experimental evidence that the edge states in nanographite disordered network govern its magnetic properties. [114, 115]. A novel edge-state-based magnetism is found in nanographite systems obtained from the heat treatment of nano-diamond particles [111].

Thermal treatment of nanographenes modifies its electronic structure which reflects on both electric and magnetic properties [115, 116]. Commercial pitch-based and phenol-based ACF show insulator-like conductivity. For these samples magnetic susceptibility obeys the Curie-Weiss law being negatively shifted at higher temperatures.

Heat treatment considerably modifies conductivity mechanisms in these systems. The in-plane size of graphitic sheets increases, and the interdomain contacts enhance, giving rise to the conduction paths between the nanographites. Conductivity successively increases, and insulating behavior changes to the variable-range hopping with a characteristic length equal to the mean size of graphene sheet. Heat treatment slightly increases the localization length, and conductivity steadily increases. Starting with some critical temperature (over 1000°C), further increase in treatment temperature leads to a sudden increase in conductivity and to the insulator-metal transition. This transition is explained in terms of creation of a percolation-path network: development of a disordered percolative metallic state.

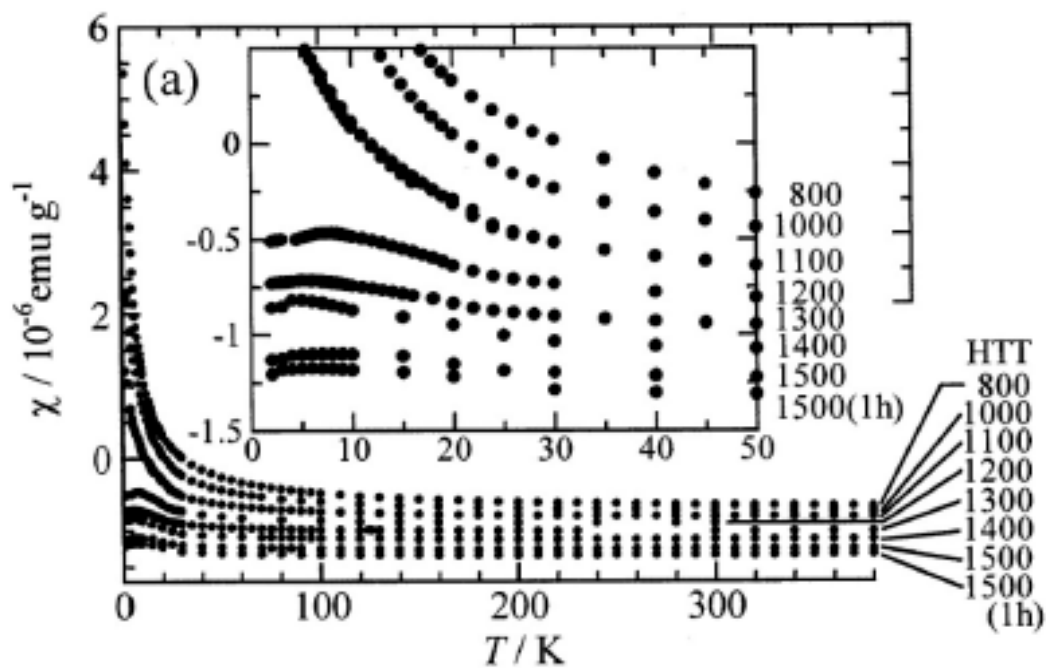

**Fig. 22 The susceptibility vs temperature plots under $H$=1 T for the ACFs heat treated at up to 1500°C. All the samples are heat treated for 15 min except 1500 (1 h) which is heat treated for 1 h. The inset shows the detailed behavior [116].**

Changes in magnetic properties patiently follow the conductivity changes (Fig. 22). Insulating conductance of ACF is accompanied by the Curie-Weiss behavior, the Weiss temperature being independent of the treatment temperature, and the spin concentration is estimated as several spins per nanographite domains. As discussed earlier, these spins have an edge origin. In a certain range of treatment temperatures the spin concentration does not alter noticeably, but the increase in temperature leads to an increase in the in-plane size, coupling of nano-graphite domains and the development of the π-electron network which is reflected in the negative shift of the susceptibility.

The further increase in temperature treatment clearly indicates that the structure is on the way to a bulk graphite: the susceptibility becomes less temperature dependent and shifts to the negative side, suggesting that the main contribution comes from orbital diamagnetism and the contribution of the edge states is reduced. The transition to a metallic behavior takes localized spins away; instead the magnetism of the itinerant electrons prevails in the metallic state.

In the vicinity of the metal-insulator transition additional features appears in the magnetic susceptibility curves: an extreme at about 7 K and a large field cooling effect below 15 K (Fig. 23), suggesting a disordered magnet state. Under the assumption that the external field varies depending on the locations of the localized spins, the authors introduce a distribution function to explain the magnetization curves. The fitting demonstrate a large deviation in the exchange field, which indicates the appearance of the spin-glass state. From the comparison of spin concentration and the domain dimensions the interspin distance is estimated as 30 Å, and it means that there is no direct interaction between the spins. The origin of this unusually long-ranged exchange interaction is discussed in terms of interactions mediated by the conduction $\pi$ - electrons with randomly distributed strengths between the localized spins.

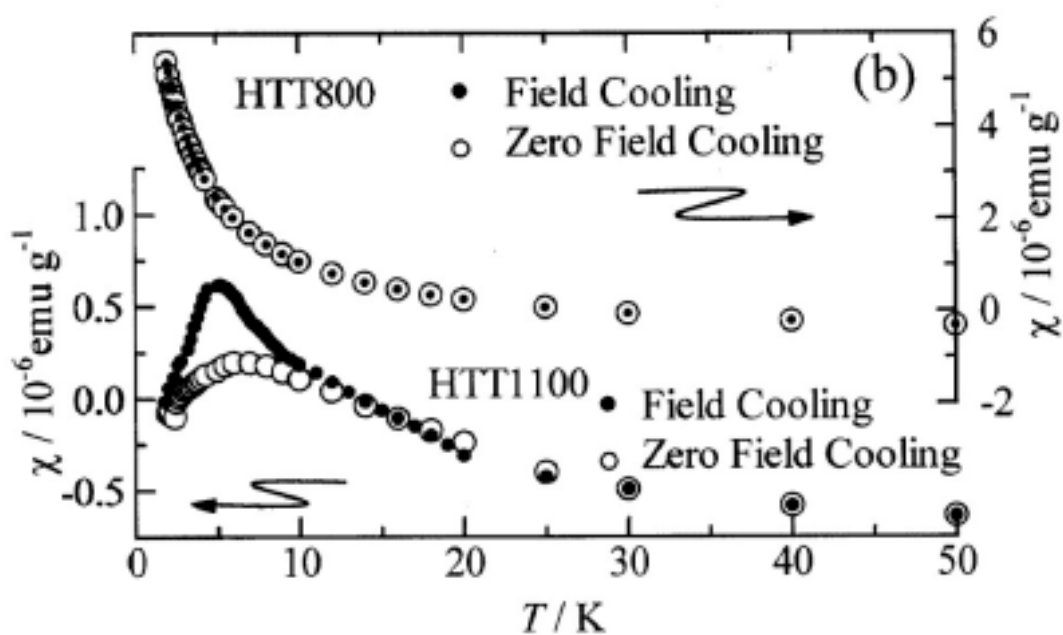

**Fig. 23. Field cooling effects on $\chi$ for high temperature treated samples, at 800 and 1100 K, where the open and solid symbols correspond to the data on zero-field and field (1 T) cooling runs, respectively. The measurements were performed in heating runs after the cooling process down to 2 K with and without field [116].**

Another source of unusual magnetic properties of graphite is the presence of topological defects. A $\pi/2$ local bond rotation in a graphitic network creates a Gaussian curvature quadrupole: two pentagons and two heptagons instead of four hexagons. [117]. Such a type of defects in graphite modifies the band structure, some of which have a flat band. This situation can lead to different electron instabilities including possible models for itinerant ferromagnetism.
 A series of one-dimensional polymers with a flat band was shown to be ferromagnetic for electron doping up to the half-filling of this band [118].
 Flat bands are expected to appear as a result of certain types of defects in nanotubes: introduction of a coronene-type defect in an armchair nanotube creates localized states, and completely flat bands appear near the Fermi level in the vicinity of the band edge [119].

## 5. Fullerene-based magnets

Magnetic properties of fullerenes are determined by the following factors: (1) fullerenes consist of carbon and (2) they contain pentagons and hexagons. Hexagonal rings contribute to a diamagnetic term, while pentagonal rings contribute to a paramagnetic term. As the paramagnetic term is nearly equal to the diamagnetic one, the fullerenes show small susceptibility values: the diamagnetic susceptibility of neutral $C_{60}$ is $- 3.4 \cdot 10^{-7}$ emu/g [120]. The paramagnetic upturn at low temperatures observed in the AC susceptibility of $C_{60}$ is usually ascribed to oxygen [121, 122], but there are indications that the positive paramagnetic term is due to the internal properties of the $C_{60}$ molecule [123].

### 5.1. $C_{60}$ charge-transfer complexes.

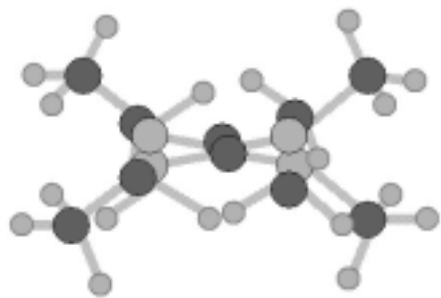

Fig. 24. A sketch of the TDAE molecule. Carbon atoms are shown as dark circles, nitrogen atoms as light circles, hydrogen atoms as small circles.

The discovery of the ferromagnetic fullerene derivative, charge-transfer salt TDAE-$C_{60}$, in 1991 attracted considerable interest, because it demonstrated that π-electron ferromagnetism at comparatively high temperatures is a reality [124]. This substance, containing only light elements, namely carbon, nitrogen and hydrogen (see Fig. 24 for the schematic structure of TDAE) has a temperature of the ordering transition towards a ferromagnetic state $T_c = 16$ K. This is more that an order of magnitude higher than $T_c = 0.65$ K for the first discovered (1990) organic ferromagnetic substance p-NPNN (p-nitrophenyl nitronyl nitroxide) [17]. There has been significant progress in studying structural, magnetic and electronic properties of this material, and we refer to the reviews "Magnetism in TDAE – $C_{60}$" [125], "Magnetism in $C_{60}$ charge-transfer complexes"[126]. The properties of fullerene derivatives, in which magnetic transitions have been reported, including the alkali-metal salts and polymers, are reviewed in Ref [127].

Here we mention only the most important features of this organic ferromagnet. TDAE, tetrakis-dimethylaminoethylene $C_2N_4(CH_3)_8$, is known as a very strong donor. Since the discovery of TDAE-$C_{60}$ ferromagnet, there has been extensive search for other charge-transfer fullerene complexes, or TDAE-compounds with other fullerenes. Magnetic signal disappears when $C_{60}$ is replaced by $C_{70}$, but the replacement of TDAE by another strong donor, cobaltocene, results in very similar magnetic properties. Ferromagnetism below 19 K in a cobaltocene-doped fullerene derivative was shown to be due to unpaired spins on fullerene molecules [128]. Several compounds showed ferromagnetic features at lower temperatures [128, 129]. There have been reports on higher $T_c$, for example TMBI-$C_{60}$ was found to have a transition at 140 K [130], but the results were not reproducible [131]. It was shown recently [132] that the molecular complex between fullerene $C_{60}$ and an organic donor BTX (9,9(')-trans-bis(telluraxanthenyl)) exhibits anomalously high magnetic susceptibility at temperatures above 130 K. The susceptibility value exceeds the one calculated under the assumption that each molecule bears one paramagnetic spin ½. From the analysis of the ESR signal, the authors make a suggestion on ferromagnetic or superparamagnetic behaviour of the samples. Magnetic properties are ascribed to the electron transfer from donor molecule BTX to $C_{60}$, which generates electron spins in the system. Anomalously high magnetism is supposed to be due to ferromagnetic correlations in the system of these spins.

The outstanding magnetic properties of TDAE-$C_{60}$ have been a subject of many scientific discussions, and still little is known about the microscopic nature of the magnetic interactions. Three questions are important: role of TDAE, role of structure, type of ferromagnetism.

There are many indications that TDAE itself is not essential for the magnetic interactions as long as other charge-transfer $C_{60}$ salts exhibit a low-temperature ferromagnetic ground state [128]. A compound consisting of two radicals, TDAE$^{+\bullet}$ and $C_{60}^{-\bullet}$, gives only one ESR line in the frequency range from 30 MHz to 240 GHz, and different mechanisms of spin cancellation were discussed.

This radical can play a certain role in the structural effects. TDAE-$C_{60}$ crystal has a monoclinic structure with unusually short distances between the buckyballs: 9.965 Å. From the ESR and conductivity studies a conclusion has been made that the exchange interactions between TDAE$^{+\bullet}$ radical and $C_{60}^{-\bullet}$ should be taken into account [133]. The charge state of TDAE is monovalent [134], and an indirect exchange via $C_{60}$ can be the reason for the absence of a TDAE ESR line [135]. The question why the coupling between $C_{60}$ and TDAE gives rise to the $g$ value close to that of $C_{60}$ remains open.

The material exists in two forms: ferromagnetic α-TDAE-$C_{60}$ and paramagnetic α′-TDAE-$C_{60}$. The paramagnetic form transforms into the ferromagnetic phase upon annealing. The α-form is a soft ferromagnet: it has a sigmoid magnetization behavior, but a very small residual magnetization, if any. The saturation magnetization is about 1 emu / g. Recent studies have shown that the mutual orientation of the neighboring $C_{60}$ molecules plays a key role in the FM exchange [136].

In α-TDAE-$C_{60}$ the fullerene molecules exist in the configurations rotated by ± 60° around the three-fold molecular axes, in addition to conventional molecular rotations. In paramagnetic α′-TDAE-$C_{60}$ phase the relative $C_{60}$ orientations are similar to those observed in other $C_{60}$ solids, and the double bond between two hexagons faces the center of the hexagon of the neighboring molecule. In the ferromagnetic phase the double bonds of one molecule faces the pentagon of its neighbor, and the $C_{60}$ molecules are ordered along the $c$ axis with the alternating orientations.

TDAE-$C_{60}$ was initially thought to be an itinerant ferromagnet. Actually, it shows an insulating behavior. Both infrared absorption spectra [137] and conductivity studies [133, 138] show that ferromagnetic α-TDAE-$C_{60}$ is a semiconductor. Temperature dependence of conductivity follows the Arrhenius law with two activation energies: 0.14 eV below $T_0 = 150$ K and 0.34 eV at higher temperatures [139]. A comparison with the infrared spectra rules out band conductivity, and two mechanisms are suggested: phonon-assisted hopping and temperature-independent tunneling. Studies of the magnetic ordering and transport properties by ESR, NMR, ac susceptibility and microwave conductivity on the α-TDAE-$C_{60}$ phase definitely exclude all models based on band ferromagnetism [133]. The present data point towards an isotropic or nearly isotropic Heisenberg ferromagnet [140].

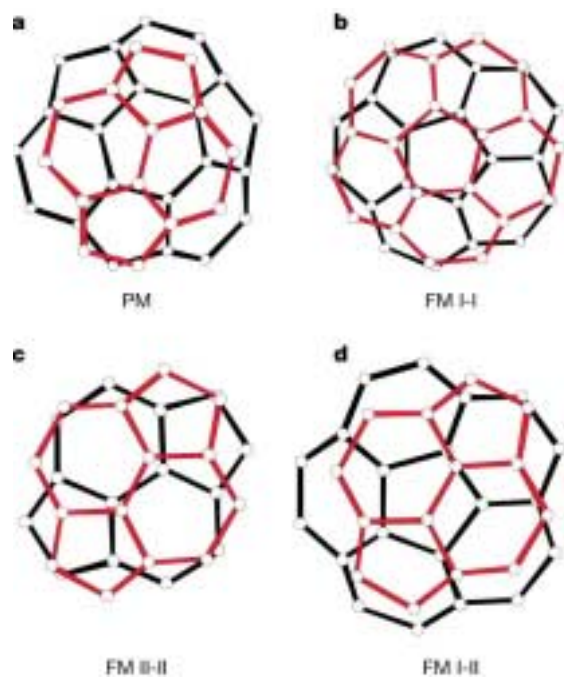

**Figure 25.** Projections of two neighboring $C_{60}$ units of TDAE − $C_{60}$ along the direction from $C_{60}$ center to $C_{60}$ center in the [001] direction. Upper molecules are shown in gray. a): The PM phase. b): The I-I configuration of $C_{60}$ molecules (FM phase). c): The II-II configuration of $C_{60}$ molecules (FM phase). d): The I-II configuration (FM phase) [136].

### 5.2. Other fullerene derivatives

There have been observations of room-temperature ferromagnetism in fullerene compounds, other than charge-transfer salts. A small increase of electron paramagnetic absorption was observed in a dimethylformamide solution of polyvinylidenefluoride in which $C_{60}$ was ultrasonically dispersed [141]. Vacuum evaporation of the solution yields a $C_{60}$-containing polymer film free from metallic contamination. This film exhibited ferromagnetic characteristics even at ambient temperatures. A transmission electron microscope observation on thin slices of the sample film showed rod-shaped aggregates of fine crystalline $C_{60}$ with cross sections of 2-8 μm which were regularly arrayed and aligned in a certain direction. By laser desorption ionization time-of-flight mass spectroscopy on the samples, $C_{60}$ adducts with R (R=H, F, $CF_3$ and polymer fragments) were detected, which seemed to be formed as $C_{60}$ radical adducts immediately after scission of the polymer chain by sonication. The observed magnetism is supposed to be due to the radical adducts represented as $C_{60}R_n$, where n is odd.

Room-temperature ferromagnetism with saturation magnetization of 0.04 emu / g was observed for palladium compounds with approximate composition $C_{60}Pd_{4.9}$ [142.] The magnetic properties decreased when the samples were stored in air. Palladium compounds, in contrast to compounds with other transition metals, are commonly diamagnetic. There is, however, interesting evidence that the structure with Pd nanoparticles sandwiched between adjacent graphite layers (without forming graphite intercalated compounds) behaves like a quasi-two-dimensional ferromagnet with very weak antiferromagnetic interplanar interaction [143]. Temperature dependencies of the field-cooled d.c. magnetization manifests a sharp deviation from the zero-field-cooled curve at 60 K for iodine-containing polymers of $C_{60}$. This can be explained by a transition to a frozen magnetic glass state [144 - 146].

The ferromagnetic properties of fullerene hydrides mentioned in the work on ultrasonically dispersed fullerenes were found in other works [147, 148]. Fullerene hydrides $C_{60}H_{18}$, $C_{60}H_{36}$, and $C_{70}H_{36}$ were studied, and magnetic measurements showed that $C_{60}H_{36}$ is a room-temperature ferromagnet with $M_s = 0.04$ emu / g. The hydrogenated fullerenes were prepared by transfer hydrogenation procedures involving 9,10 – dihydroantracene. Samples produced from several batches showed a similar dependence of magnetization versus external field, whereas other compositions preserved a diamagnetic character. The authors supposed that ferromagnetism is caused by the peculiarities of the hydride structure and is of a kinematic nature: differences in electron acceptor properties for pentagons and hexagons could be modified by the hydrogen-carbon bonds [149].

Further investigations of hydrofullerites confirmed the initial observations [150]. The samples were prepared in a different way, under pressure of 0.6 and 3 GPa and at the temperatures 250 – 350 °C in an excess of hydrogen, and twelve $C_{60}H_x$ samples with $x$ varying from 24 to 32 were obtained. The hydrofullerites have either a *fcc* or *bcc* lattice formed of $C_{60}H_x$ units. Every sample shows a hysteresis in its magnetization curve with the coercivity about 100 Oe. Most hydrofullerites have low values of saturation magnetization, about 0.001 – 0.003 $\mu_B$ / $C_{60}$. However, three samples showed rather big values of magnetization: 0.046, 0.054 and 0.16 $\mu_B$ / $C_{60}$. All three samples were synthesized under the same pressure-temperature conditions, and had the *fcc* structure with the composition $C_{60}H_{24}$.

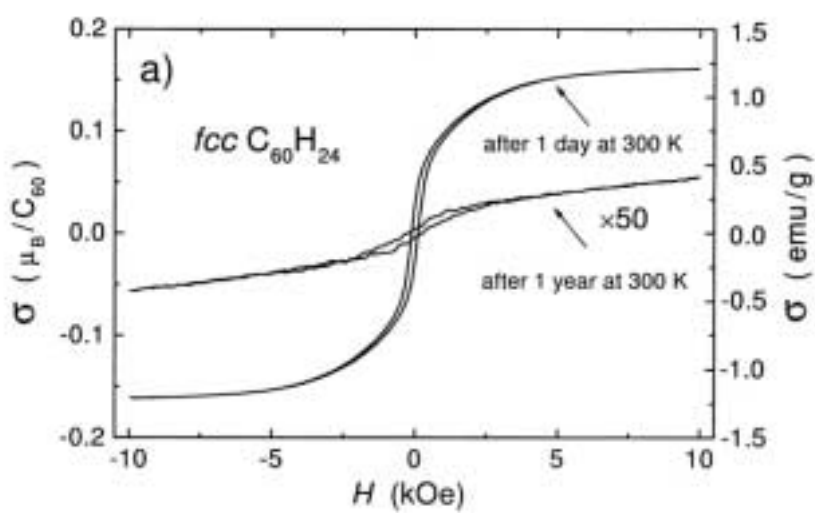

**Figure 26. Magnetization σ as a function of magnetic field $H$ at room temperature for the $C_{60}H_{24}$ sample synthesized at hydrogen pressure of 0.6 GPa and T = 350°C and exposed to ambient conditions for 1 day and for 1 year [150].**

The samples were subjected to the atomic-emission analysis for the metal impurities. The concentration of all detected metals in the sample with magnetization of 0.16 $\mu_B$ / $C_{60}$ (1.2 emu / g) is as follows (Wt.%). Fe: 0.01; Ni: 0.002; Pd: 0.01; Al: 0.05; Cu: 0.1. The contribution of metallic impurities can lead to the two orders of magnitude less magnetization. It is clear that magnetic ordering is an intrinsic property of hydrofullerites.

A circumstantial evidence of the intrinsic nature of ferromagnetism is an aging phenomenon. The storage of the samples at ambient conditions results in a drastic decrease of the magnetization (Fig. 26). A 1-year storage brings the samples to a diamagnetic state without noticeable changes in their composition and lattice parameter.

The $\sigma$ (H) curves are invariant in the temperature range 80 – 300 K, showing that the Curie temperature lies well above the room temperature for all samples including the ones stored in air. The combination of high Curie temperature and small values of magnetization shows that the sample is unlikely to be a bulk collinear ferromagnet. Small $\sigma$ and high $T_c$ can be explained from the viewpoint of spin canting or defects in an antiferromagnetic structure. This explanation is poorly adapted to aging of the samples with the transition to diamagnetic behavior: after the disappearance of spin alignment the sample should remain antiferromagnetic. Another possibility is that the samples comprise a mixture of diamagnetic and ferromagnetic phases: this is consistent with both aging phenomenon and the scattered values of magnetization of as-prepared samples.

### 5.3. Photopolymerized fullerenes

It is possible to observe ferromagnetic properties of $C_{60}$ without adding strong donors like TDAE and without treatment with iodine or hydrogen. Exposure of the $C_{60}$ crystals to light in the presence of oxygen leads to the appearance of saturating behavior in the magnetic field with an obvious hysteresis loop [151].

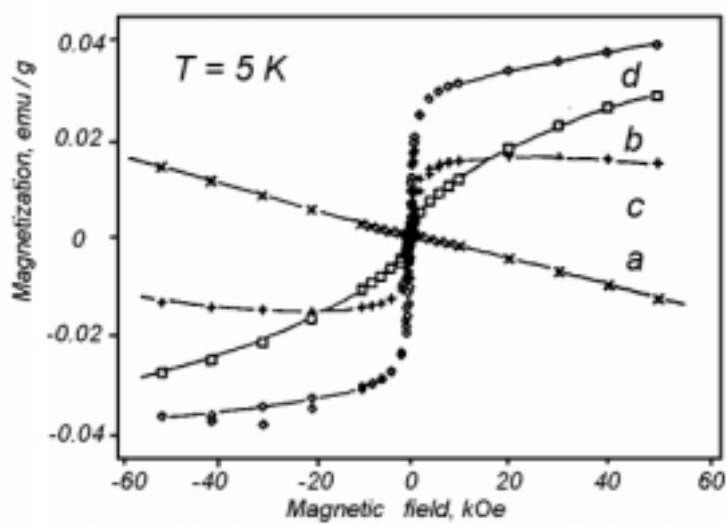

**Figure 27. The field dependence of magnetization at T = 5 K for (a) the pristine $C_{60}$ crystal, (b) the sample exposed to light in oxygen for 2.5 hours (c) baked at 400°C for 2.5 hours, and (d) exposed in air for 3 months [151].**

Pristine van der Waals $C_{60}$ crystal is a diamagnetic (Fig. 27, curve a). The susceptibility value for the samples stored in darkness and in vacuum is practically independent of temperature. The crystals are extremely sensitive to oxygen, which quickly penetrates into the bulk. The paramagnetic upturn at low temperatures is always observed in the AC susceptibility of pristine $C_{60}$ except in the case of specially prepared $C_{60}$ single crystals, never exposed to oxygen.

Different situation occurs if the sample is exposed to oxygen under the action of the visible light. The susceptibility changes its sign to positive in the whole temperature range, and its absolute value progressively increases. Exposure during 2.5 hours brings noticeable features of ferromagnetism: non-linear magnetization process at low fields (Fig. 27, curve b), and magnetization increases with the increase of the exposure time (Fig. 27, curve d). Saturation value at high fields is $1.4 \cdot 10^{-2}$ emu / g, remanent magnetization is about 10% of the saturated value, coercive force 100 Oe.

It is known that the physisorped oxygen can be driven away from the $C_{60}$ crystal by heating in vacuum. Heating of the sample, which was oxygen-exposed under the action of visible light, does not restore it to a pristine state. This sample still keeps and even increases the ferromagnetic behavior, but the paramagnetic background changes to diamagnetic one (Fig. 27, curve c).

The most striking feature of the magnetization curves is that they remain practically the same from the temperature of 5 K (shown in Fig. 27) to the room temperature. Above 300 K magnetization starts to decrease slowly, but remains finite up to 800 K.

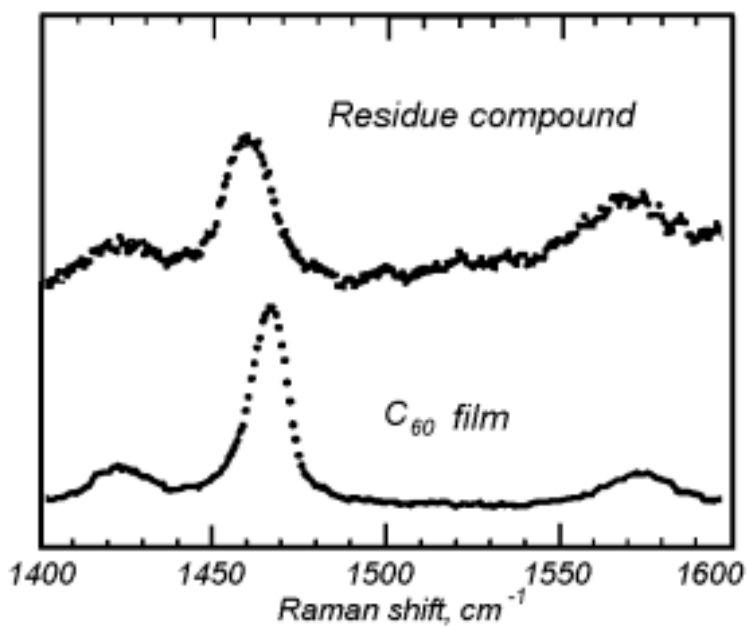

**Fig.28. The Raman spectra of the ferromagnetic residue compound and the pure $C_{60}$ film [151].**

The saturation value $M_s = 1.4 \cdot 10^{-2}$ emu / g corresponds to approximately 0.001 $\mu_B$ / carbon atom. This small value was, however, increased by separation of the material into a magnetic and a non-magnetic part. Separation was made by dissolving in toluene, a solvent for pristine $C_{60}$. The undissolved part showed the same ferromagnetic behavior, both qualitatively and quantitatively, and can be considered as a concentrated ferromagnetic phase. This residue substance has a hundred times larger magnetization: 0.1 $\mu_B$ / carbon atom.

The x-ray diffraction measurements reveal that the position of the main peaks are very close to those for a $C_{60}$ crystal, but the peaks are sufficiently broadened. Raman spectra in the vicinity of the pentagonal pinch mode show that this mode, originally peaking at 1468 $cm^{-1}$, is shifted to 1458 $cm^{-1}$. This is typically observed for the orthorhombic phase of polymerized $C_{60}$: linear polymerization.

The experiments described in Ref. [151] were repeated by the author of this paper [152] without knowing about Y. Murakami and H. Suematsu's work. We have obtained very similar results. Measurements of the AC susceptibility were made on the commercial $C_{60}$ produced by the Term USA. The material was stored in darkness under a dynamic vacuum of $10^{-1}$ Torr; this vacuum is not enough to keep the material from oxygen. The susceptibility of the pristine fullerene powder at temperatures above 6 K is negative and almost temperature independent. At low temperatures the susceptibility increases (decreases in the absolute value) and even changes its sign, showing weak paramagnetism at 2 – 4 K due to the superimposed Curie-like term, which is most likely due to oxygen contamination. In our experiments the measured value of the room temperature susceptibility is − 3.1·$10^{-7}$ emu/g Oe, which shows the relative purity of our samples: the amount of paramagnetic impurities is low. In contrast to pristine $C_{60}$, the magnetization of light-and-air exposed $C_{60}$ is positive, but it shows a similar upturn at the same temperature. The magnitude of the susceptibility increases with increasing exposure time. When allowance is made for the superimposed Curie term, the susceptibility of light-and-air exposed fullerenes increases slightly with temperature.

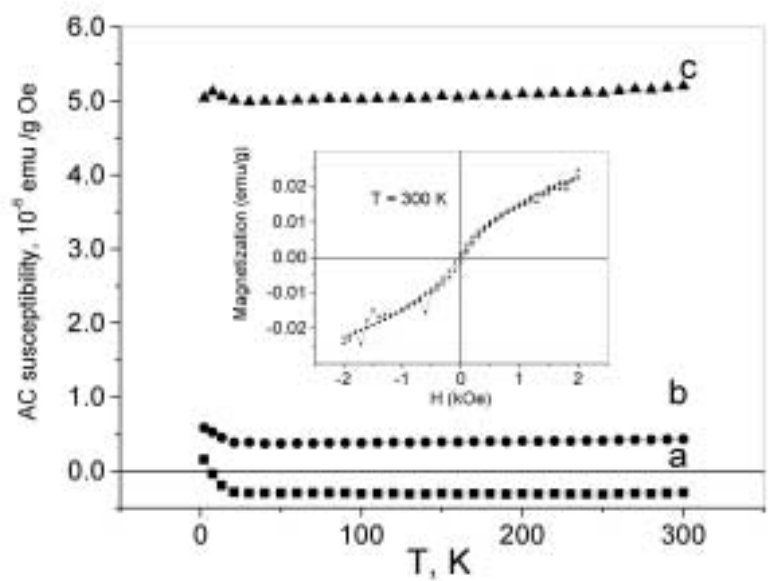

**Fig 29. Temperature dependencies of AC susceptibility for (a) commercial $C_{60}$ powder stored for a year in darkness under low vacuum; (b): exposed to light in air for 24 hours; (c) exposed to light in air for one month. Inset: magnetization curve for the sample (c) at room temperature.**

A non-linear magnetization curve is observed only for the fullerenes exposed to light in the presence of oxygen; without light only an enhancement of the paramagnetic Curie-term is registered. It is well known, that oxygen-free $C_{60}$ films and crystals are transformed into a photopolymerized state under the action of UV and visible light [153]. When $C_{60}$ is simultaneously exposed to oxygen and UV or visible light, the vibrational and optical absorption spectra testify to two processes: (i) a photoassisted diffusion of molecular oxygen into solid $C_{60}$ and (ii) an oxidation of $C_{60}$ [154]. The photopolymerization reaction occurs between the monomer $C_{60}$ in the excited (triplet) state and another monomer in the ground state. It is commonly believed that the polymerization does not occur in the presence of oxygen, because oxygen quenches the $C_{60}$ triplet state [155]. Our experiments show, however, that the physical properties of $C_{60}$ films exposed to the laser light in air and in the vacuum of $10^{-6}$ Torr are very similar, and even the time required to obtain a toluene-insoluble phase is the same [156]. In both cases the laser exposure is followed by a broadening and softening of the Raman-active $A_g(2)$ mode which shifts from 1468 to 1458 cm$^{-1}$. The exposed samples were treated with standard organic solvents attacking the unpolymerized phase of the fullerite. The resulting films consist of voids (where the $C_{60}$ is washed out), carbon and oxygen, but the air-treated films contain 48% of voids, whereas the oxygen-treated films are more dense, contain 30% of voids, and have an average stoichiometry of $C_{60}O_{13.8}$. We thus argue that the light-and-air exposed $C_{60}$ is a photopolymer, containing oxidized $C_{60}$ molecules.

A comparison of the Raman spectra for the samples exposed to light and air during 24 hours and one month (the same samples as (b) and (c) of Fig. 29) clearly shows that the stronger the magnetic properties, the higher the 1458 / 1468 ratio for the satellites of the $A_g(2)$ mode, corresponding to higher fraction of the polymerized phase (Fig. 30).

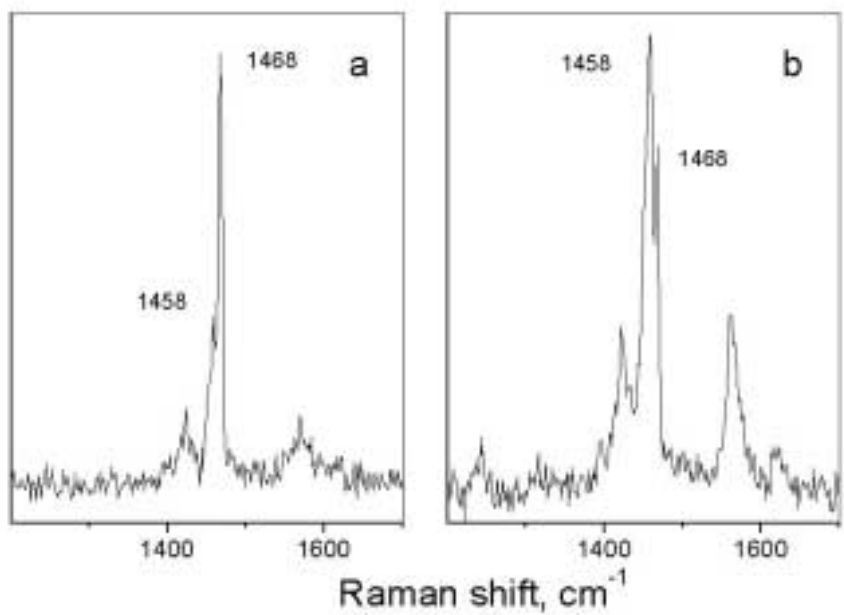

**Fig. 30. Raman spectra for $C_{60}$ exposed to light in air for 24 hours (a) and for one month (b). Saturation magnetization for sample (b) is 10 times higher.**

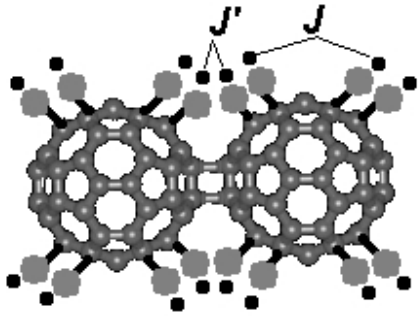

**Fig. 31.** A schematic model for the ferromagnetic interaction between oxygen radicals (gray circles). Adapted from [151].

Y. Murakami and H. Suematsu suggest a model, which takes into account both polymerization and oxidation of $C_{60}$. They regard oxidation of $C_{60}$ molecules as a mechanism of unpaired spin formation: the O2 molecule dissolves by light irradiation to make a carbon-oxygen pair, which has an unpaired electron. The spins could participate in the exchange interaction, and for the spins on neighboring molecules (J´) the interaction is possibly stronger than for the spins on the same molecule (J).

### 5.4. Pressure-polymerized fullerenes.

It is possible to form all-carbon polymers by subjecting fullerenes to high pressure-high-temperature treatment. The first fullerene polymer obtained under pressure had a rhombohedral (Rh) structure [157]. In the subsequent experiments, $C_{60}$ was found to transform upon heating under pressure into three different phases: orthorhombic (O), tetragonal (T) and rhombohedral (Rh) [158]. The combination of pressure and temperature, which is necessary to obtain the desired phase, has been studied by several groups, and for the details we refer to the review [159].

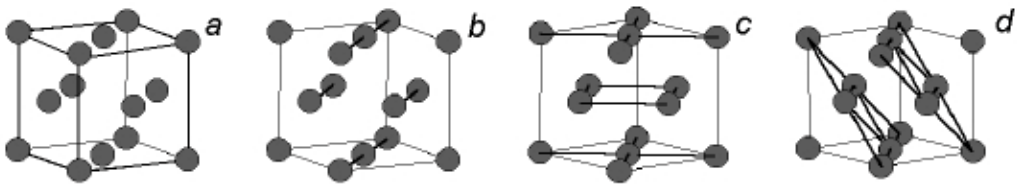

**Fig. 32.** Transformations of fullerene $C_{60}$ under pressure: (a): pristine *fcc* lattice, (b): orthorhombic phase, (c): tetragonal phase, (d): rhombohedral phase.

Figure illustrates the formation of different polymerized phases from the pristine face centered cubic $C_{60}$ lattice. Applying pressure at comparatively low temperatures (less than 500 K) yields the orthorhombic (O) phase. High values of pressure and temperature (p > 5 GPa, T > 600 K) result in the rhombohedral phase. In the intermediate pressure-temperature range, a mixture of tetragonal and rhombohedral phases is usually formed, but it is possible to obtain a pure tetragonal phase by manipulating the preparation conditions in a preassigned order.

The presence of a magnetically ordered phase was revealed in the Rh-$C_{60}$ prepared at a pressure of 6 GPa and in a narrow temperature range, slightly higher than 1000 K. The maximum magnetization value was achieved at 1075 K [160]. Magnetisation loops measured in the field range - 2 kOe < H < 2 kOe, for temperatures 10 K and 300 K show nearly identical hysteresis with $M_r$ = 0.015 emu/g and $H_c$ = 300 Oe. A saturation of the magnetisation is clearly seen above ~ $2 \cdot 10^4$ Oe. Using the spin concentration value obtained from the electron spin resonance data, n = $5 \cdot 10^{18}$ cm$^{-3}$, we estimate the magnetic moment as 0.4 $\mu_B$ per electron. From the temperature dependence of the magnetisation at a fixed field of 0.2 T, and also from the temperature dependence of the remanent magnetisation obtained at H = 0 after decreasing the applied field from 2000 Oe, the Curie temperature is estimated as 500 K (Fig. 33).

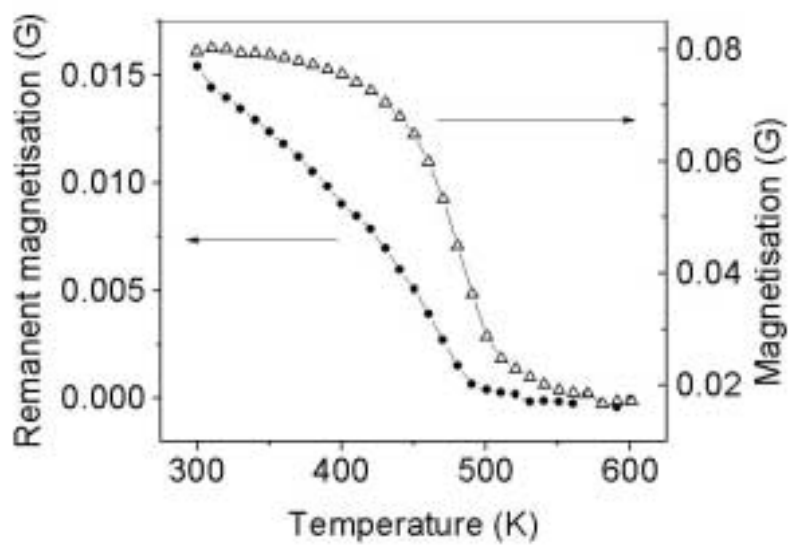

Fig. 33. Magnetisation of the Rh-$C_{60}$ in a fixed applied field of 0.2 T (upper curve, triangles) and the remanent magnetisation obtained at H = 0 (lower curve, circles) as a function of temperature.

In order to elucidate the nature of the magnetic properties of pressure-polymerized fullerenes, different conditions of sample preparation have been tried. Experiments were performed at the pressure of 2.5 GPa, which is favorable for the formation of the tetragonal phase. We have obtained precisely the same result: the magnetic phase is formed in a very narrow temperature range with the maximum at 1025 K. The ferromagnetic behavior totally disappears for samples prepared at 1100 K and higher, and the susceptibility reverts to a diamagnetic behavior. Another observation is that the magnetic properties are very sensitive to preparation time, and the search for optimal preparation conditions is to be made in at least 3-dimensional p-T-t space.

An increase in pressure to 9 GPa results in the following: for temperatures as low as 800 K the polymerized samples are paramagnetic (prepared from the pristine diamagnetic material). For the temperature of 900 K a distinct hysteresis is superimposed over the paramagnetic signal. Further increase in temperature leads to an abrupt transition to the diamagnetic behavior (Fig.34) [161]. The reduction of the ferromagnetic character and the return to a diamagnetic behavior indicates the considerable loss of magnetic centres during graphitization. A small ferromagnetic contribution is seen even after the cage collapse, but this feature weakens as the preparation temperature increases. The variation in remanence magnetization with temperature in comparison with the hardness variation is shown in Fig. 35: the remanence shows a dramatic increase at 800 K which is before the temperature of $C_{60}$ cage collapse [161].

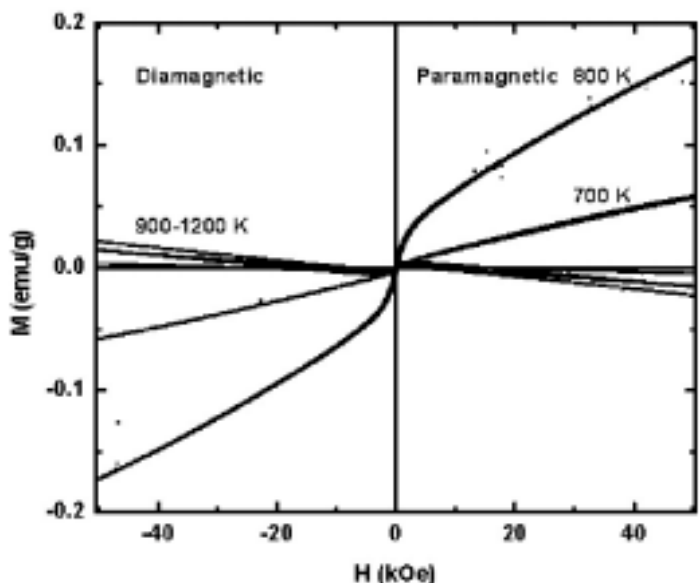

**Fig. 34.** M versus H loops of high T/p treated $C_{60}$ at 9 GPa over the range 700–1200 K [161].

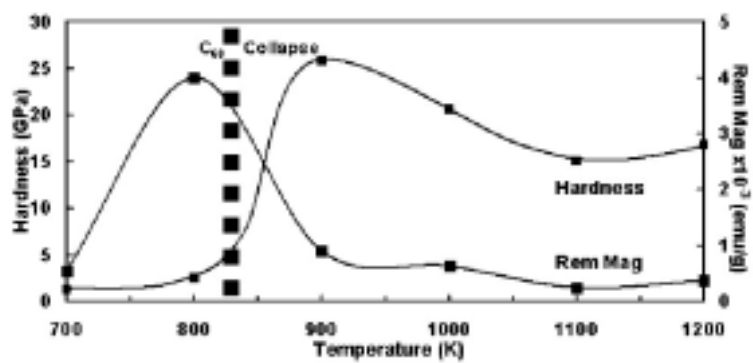

**Fig. 35. Hardness and remanent magnetization graph for samples treated at 9 GPa, indicating the region of radical centre generation [161].**

These observations suggest that the type of polymerization is probably not critical for the magnetic properties, because the magnetic phase invariably appears on the boundary between polymerized fullerenes and postfullerene phase [162]. The three points at which the magnetic properties were found to be most pronounced: (2.5 GPa, 1025 K; 6 GPa, 1075 K; 9 GPa, 900 K) are situated at the critical points of the pressure-temperature plane showing the various phases of $C_{60}$ created under different conditions.

### 5.5. Origin of fullerene ferromagnetism

Fullerenes differs from graphite in being defect-free. This circumstance governs the choice of models for fullerene magnetism.

*5.5.1. HTFLCC*

Pure polymerized fullerenes, oxygenated fullerene polymers and hydrofullerites share a common trait: low magnetization values apparently contradict to the high temperatures of the magnetic ordering. The ferromagnetic properties arise in a very narrow region of preparation conditions. There is a temptation to draw a parallel between unusual magnetism of fullerenes and hexaborides. A phenomenon of high-temperature weak ferromagnetism at low-carrier concentration (HTFLCC) with no atomic localized moments has recently been found in alkaline-earth hexaborides $CaB_6$, $SrB_6$, $BaB_6$. Lanthanum or thorium doped divalent hexaborides bear a weak ferromagnetic moment, Ca-deficient $Ca_{1-\delta}B_6$ or isovalent substitutional alloy $Ca_{0.995}Ba_{0.05}B_6$ are ferromagnetic, and this phenomenon was found even in nominally pure hexaborides.

Boron is a next-door neighbor of carbon in a Periodic table. Hexaborides are binary compounds with a simple cubic structure and a large variety of physical properties: for example, $EuB_6$ is a ferromagnet, exhibiting a colossal magnetoresistance; $LaB_6$ is a low-temperature semiconductor. Alkaline-earth hexaborides are believed to be either semiconductors [163], or semimetals [164], in close vicinity to the border between semimetals and small-gap semiconductors and having a peculiar configuration of the electronic excitation spectrum: the valence and conduction band are separated with a gap in all points of the Brillouin zone, except for the X points, where a week overlap does exist. Low electron doping with $La^{3+}$ (order of 0.1%) was shown to result in an itinerant type of ferromagnetism stable up to 600 K [165] and 900 K[166]. Saturation moment is quite sensitive to the doping level and reaches 0.07 $\mu_B$ / electron, the electron density being $7 \cdot 10^{19}$ $cm^{-3}$. Sharp decrease in saturation with the increase of doping level rules out the effects of an accident contamination, but requires consideration from the viewpoint of the electronic band structure. Taking into account the absence of localized magnetic moments, magnetic order was ascribed to the itinerant charge carriers: a ferromagnetic phase of a dilute three-dimensional electron gas [167].

A different approach is based on the formation of excitons made of electrons and holes in the overlap region around the X point (Fig, 36, a) [168]. Electrons and holes are created as a result of this overlap. Coulomb attraction between these electrons and holes can lead to a condensation of the bound exciton pairs. Condensation opens a gap in the quasiparticle spectrum (Fig. 36. b). Excitonic insulator is thus created, which contains a condensate of a spin-triplet state of electron–hole pairs. A ferromagnet with a small magnetic moment but with a high Curie temperature can be obtained by doping an excitonic insulator. Doping provides electrons to the conduction band, change the electron-hole equilibrium, paring becomes less favorable, and extra electrons become spatially aligned. The alignment can be understood from the following viewpoint: the spin distribution for the extra carriers is asymmetrical, because the asymmetry preserves the most favourable paring condition for one spin orientation. The quasiparticles carry only a fraction of the Bohr magneton: calculations yield the maximum value of 0.14 $\mu_B$ per doped electron at the doping level of 0.35%. Higher concentration leads to higher values of the Curie temperature, but smaller saturation moment. Proximity of a single graphene sheet to the electronic insulator phase was proposed as one of possible origins of weak ferromagnetism in bulk graphite [169].

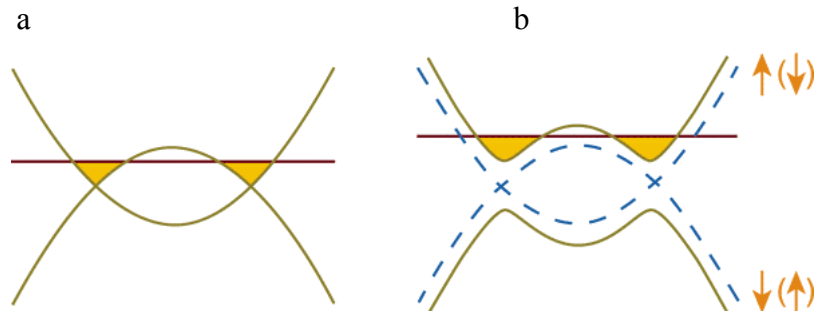

**Fig. 36. Structure of energy spectrum in a semimetal (left) and in an excitonic insulator (right). States in shaded volumes are occupied by doped electrons. Arrows indicate electron and hole spins for the case of weak ferromagnetism [168].**

The idea of ferromagnetic instability in the excitonic metal was further developed by adding the effect of imperfect nesting on the excitonic state [170]. One more idea is based on the exciton condensation model: $CaB_6$ is an antiferromagnet, but the ferromagnetism is induced by the magnetoelectric effect, and the doping works mainly as a source of the electric field [171]. Magnetoelectric effects can be caused by not only by doping, but also by introducing surfaces, and should be best observed in thin films. ESR experiments give some evidence that ferromagnetism of hexaborides is not a bulk phenomenon, but spins exist only within the surface layer of approximately 1.5 μm thick [172].

The model of an excitonic insulator requires the cubic symmetry with the presence of three equivalent valleys in the energy band. Weak ferromagnetism, however, was observed in a compound $CaB_2C_2$, which crystallizes in a tetragonal structure without the threefold band degeneracy [173]. The magnitude of the ordered moment is $10^{-4}$ $\mu_B$ per formula unit, and the transition temperature is 770 K. The authors consider peculiar molecular orbitals near the Fermi level to be crucial to the high-$T_c$ ferromagnetism. Calculations of the HOMO and LUMO structures show remarkable similarities for $CaB_2C_2$ and hexaborides.

Later observations of anomalous NMR spin-lattice relaxation for hexaborides showed the presence of a band with a coexistence of weakly interacting localized and extended electronic states [174]. Measurements of thermoelectric power and the thermal conductivity in the range of 5 – 300 K can be described by scattering of electrons on acoustic phonons and ionized impurities in the conduction band, which is well separated from the valence band [175]. The explanation of the origin of weak but stable ferromagnetism with the model of a band overlap becomes a problem.

The starting point of the majority of models is a semimetallic structure with a small overlap. However, recent parameter-free calculations of the single-particle excitation spectrum based on the so-called GW approximation reported the formation of a rather large band gap: $CaB_6$ is not a semimetal but a semiconductor with a band gap of 0.8 eV [163]. Angle-resolved photoemission answered fundamental question of whether divalent hexaborides are intrinsic semimetals or defect-doped bandgap insulators: there is a gap between the valence and conduction bands in the X point, which exceeds 1 eV [176]. Assuming that $CaB_6$ is a semiconductor, magnetism is considered to be due to a La-induced impurity band, arising on the metallic side of the Mott transition for the impurity band [163].

Magnetic properties of these structures can be considered from the point of view of imperfections in the hexaboride lattice. It was found that of all intrinsic point defects, the $B_6$ vacancy bears a magnetic moment of 0.04 $\mu_B$; an explanation for ordering of these moments can be obtained from the assumption that in the presence of compensating cation vacancies a $B_6$ vacancy cannot be neutral [177]. A support for the impurity-band mechanism can be found from the electrical and magnetic measurements on several La-doped samples [178]. All the samples show metallic behavior of conductivity. Prepared at nominally identical conditions, some of the samples are paramagnetic, and some are ferromagnetic, suggesting that ferromagnetic state can be spatially inhomogeneous. On the other hand, the models for a doped excitonic insulator also include spatial inhomogeneity [179] and phase separation with appearance of a superstructure [180].

*5.5.2. Fullerenes in the vicinity of MIT*

The similarity between fullerenes and hexaborides is not only in high Curie temperature and low magnetization values, but also in that the ferromagnetic properties arise in the close vicinity to the metal-dielectric transition.

What happens to the fullerenes on their way from spherical cages to disordered graphite-like structures? Answering the above question would be a key to clarify the mechanism of the ferromagnetism.

The band structure of two-dimensional (rhombohedral and tetragonal) fullerenes was calculated in the local density approximation. The electronic structure of these structures was emphasized to have a 3D nature because of the small interlayer spacing and strong interaction between the layers.

The following processes occur during polymerization and strongly affect the electronic structure of the polymer [181, 182]. Eight (T phase) or twelve (Rh phase) $sp^3$-like carbon atoms appear at each molecule as a result of interfullerene bond formation. The $\pi$-electron system looses its spherical character, and atoms retaining $\pi$ state are separated in two groups: those lying above and below the polymerization plane. The intermolecular distances become shorter not only within the polymerization plane, but between the planes as well. Fullerene molecules are distorted, their diameter in the direction normal to the polymerization plane decreases.

Although polymerization transforms $sp^2$ into $sp^3$ (diamond-like) bonds, the energy gap does not approach that of diamond; by contrast it becomes narrower (Fig. 37). The calculated values for Rh phase is 0.35 eV, for T phase is 0.72 eV, whereas the bandgap for the pristine non-polymerized $C_{60}$, calculated by the same method, amounts to 1.06 eV. (The theoretical values for the $C_{60}$ bandgap vary between 1.5 and 2.4, the most reliable experimental data give 2.3 eV [183]). In the case of molecular $C_{60}$, the triply degenerate states form an isolated conduction band. For both two-dimensional polymers the lower branch of the conduction band is separated from the higher states. The dispersion of the band is significantly stronger than in the *fcc* $C_{60}$ case. Strong dispersion is also observed deep inside the valence band related to the σ states, due to the formation of the intercluster bonds. The strong changes in electronic structure are mainly due to the fact that the features of the lower conduction and higher valence bands are determined by the topology of the π-electron system. In case of two-dimensional polymers only 48 (Rh phase) or 52 atoms (T phase) of 60 are in the π state. The charge density in the interfullerene bond is the same that for intrafullerene bonds, but the charge density distribution in the perpendicular direction drastically differs from that within the plane, and no noticeable charge density is present in the interplanar space. Thus, the interlayer interaction in 2D $C_{60}$ polymer is considered to be of the van der Waals type, similar to graphite.

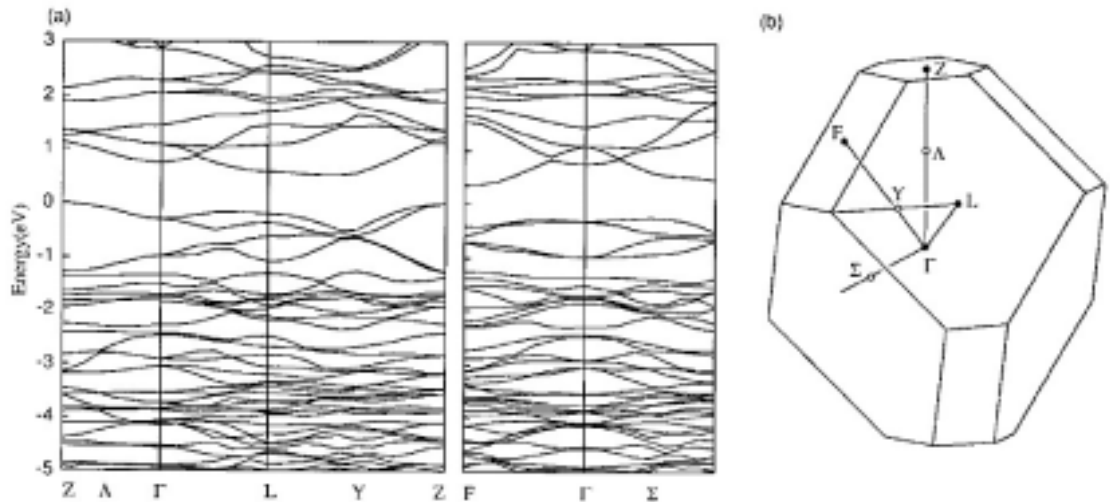

**Fig. 37. (a):** Band structure of the two-dimensionally polymerized rhombohedral $C_{60}$. Energy is measured from the top of the valence band at the Z point. **(b):** Symmetry points and lines in the first Brillouin zone of the rhombohedral lattice. [181]

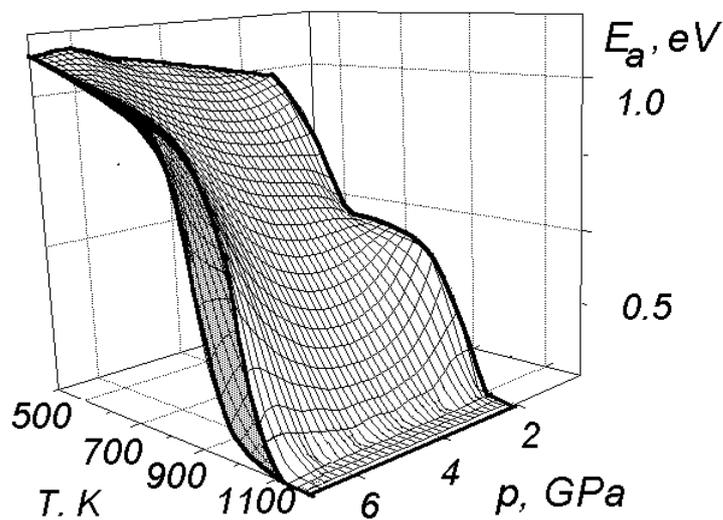

**Fig. 38.** The activation energy $E_a$ for polymerized $C_{60}$, deduced from the DC conductivity, as a function of polymerization conditions.

Calculations indicate, that in contrast to 2D polymerized fullerenes, the electronic structure of the 3D polymer is metal-like [184]. The metal-insulator transition is predicted to occur at the pressure of 20 GPa. The metallic state is created through the complete change of the π-electronic system, resulting from the distortion of the molecules and overlapping the π-like states of the triply coordinated atoms with the hybridization essentially different from $sp^2$.

In the experiments we have observed the metal-insulator transition, which apparently has a different origin

Pristine $C_{60}$ is a semiconductor, and this is reflected in its electrical and optical properties. We have studied the changes in the conductivity activation energy ($E_a$) for fullerenes compressed at different pressures $p$ and temperatures $T$. An increase in treatment temperature at a certain pressure always leads to a decrease in the activation energy. Fig. 36 demonstrates not the one-to-one correspondence between $p$, $T$ and $E_a$, but the general trend, because each point in the $p$-$T$ diagram is characterized by a set of activation energies interconnected by the Meyer-Neldel rule for disordered media [185]. A common feature for the whole pressure region is that at temperatures above 1100 K the activation energy drops to zero due to the collapse of fullerene cages. Destroyed fullerenes behave as a poor metal.

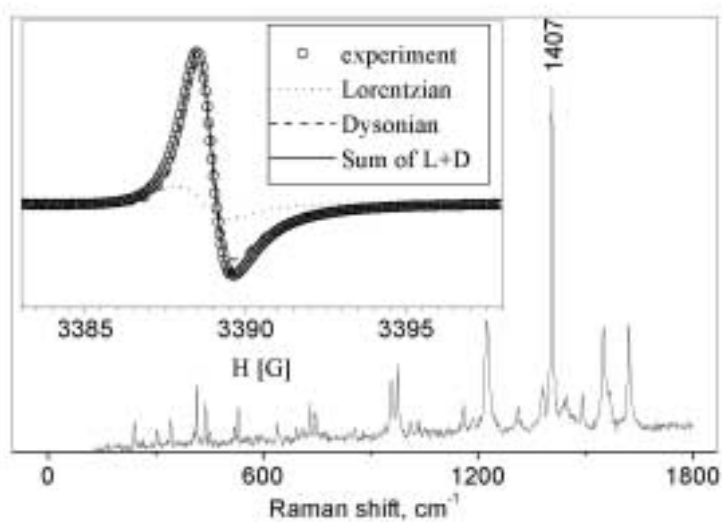

**Fig. 39. Raman spectrum for the rhombohedral phase of polymerized $C_{60}$. Inset: ESR experimental results fitted as the sum of Lorentzian and Dysonian contributions.**

The features of metallic behaviour of fullerenes appear long before the structural and optical methods detect the presence of destroyed cages. Pressure-polymerized samples prepared at temperatures less than 900 K, show narrow resonance with a symmetrical line. The spin concentration and the g-factor are almost independent of treatment pressure and temperature, being $2 - 7 \cdot 10^{17}$ g$^{-1}$ and 2.0024 – 2.0027, respectively. When the temperature increases, the lineshapes acquire the characteristic asymmetry of spin resonance in metals.

Fig. 39 shows the Raman spectrum for the sample prepared at 975 K: it manifests the pure rhombohedral $C_{60}$ phase. Inset shows the ESR experimental results fitted as the sum of Lorentzian and Dysonian contributions. The A/B ratio between the maximum and minimum of the derivative of the absorption versus field is equal to 2.45. Mechanical grinding restores the symmetrical ESR shape, proving that the effect is due to the metallic conductivity.

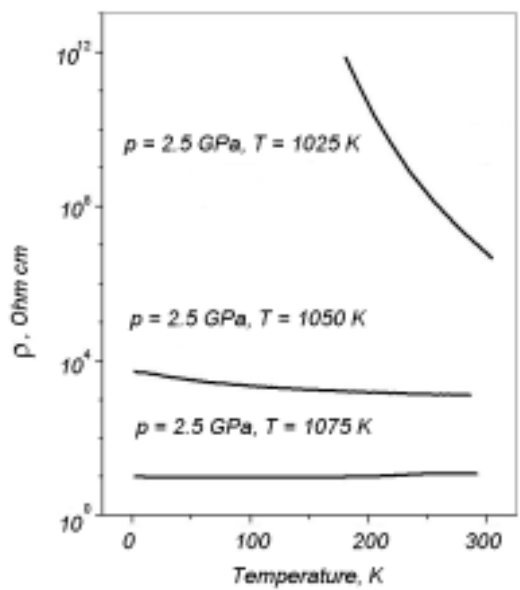

**Fig. 40 Temperature dependencies of resistivity for the tetragonal $C_{60}$ polymers near the MIT.**

Fig. 40 shows the changes in transport properties of polymerized fullerenes, prepared at 2.5 GPa (tetragonal phase). The most pronounced ferromagnetic properties are observed for the sample prepared at 1025 K, whereas the sample prepared at 1075 K has already undergone the transition to a diamagnetic behavior.

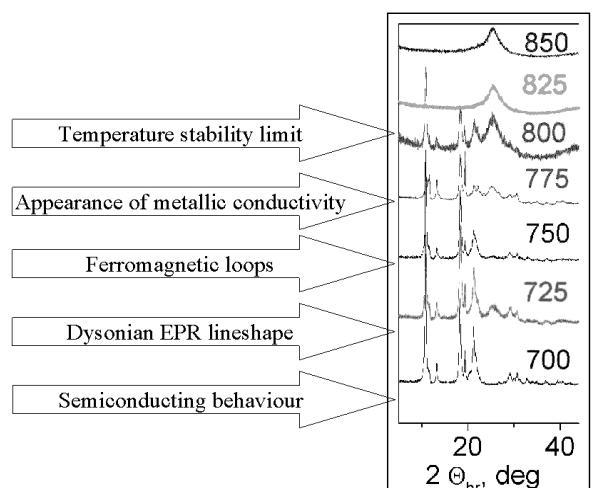

**Fig. 41. Correlation between electrical, magnetic and structural properties for the fullerenes polymerized at 6 GPa and indicated temperatures (°C).**

Temperature dependencies of conductivity for the rhombohedral phase are similar, but much more complicated, because the structure is highly oriented, and conductivity is anisotropic [186]. A general trend for the changes in electrical and magnetic properties of the rhombohedral phase of polymerized $C_{60}$ is shown in Fig. 41. This Figure illustrates also the structural changes detected by X-ray diffraction. The samples synthesized at p = 6 GPa and temperatures in the range 1000 – 1075 K reveal a gradual transition from the semiconductor to semi-metallic behavior. Samples prepared at higher temperatures exhibit great anisotropy in resistivity. The main features of the behaviour of these samples are: (i) in the z-direction the resistivity is of the order of several kOhm cm and decreases with the temperature; (ii) in the x-y direction resistivity is less than 1 Ohm cm and shows a minimum at a certain temperature. The transition to a pronounced metallic behavior is detected only for the totally destroyed fullerene phase.

*5.5.3 Models for ferromagnetism of pressure-polymerized fullerenes.*

The complete analysis of the experimental data on pressure-polymerized fullerenes shows that the ferromagnetic behavior appears quite close to the boundary where the fullerenes are destroyed. However, when sufficient amorphous-like carbon is mixed with fullerene molecules, that its presence is detectable by Raman and X-ray techniques, the ferromagnetic properties quickly decay. Starting with this evidence, the following scenario can be suggested.

One assumption is self-doping: a negligible amount of fullerene cages is already destroyed under these conditions. The electrons coming from the broken cages would enhance the conductivity by several orders of magnitude, as is observed in our experiments. Thus a situation is created, similar to that considered in p. 5.5.1.

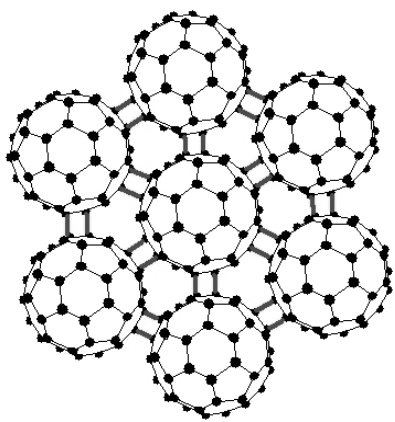

**Fig. 42 . A model of rhombohedral phase of polymerized $C_{60}$**

R. Wood [161] disproves this explanation showing that the radical centres are formed in the polymeric state before collapse, without damage to the buckyballs. The TEM images show that the samples are polymeric and crystalline in nature (Fig. 43). We note, however, that the amount of wracked fullerenes can be negligibly small and undetectable in structural studies. Two conditions are necessary for fullerene network to become ferromagnetic: highly-oriented structure and the presence of unpaired spins.

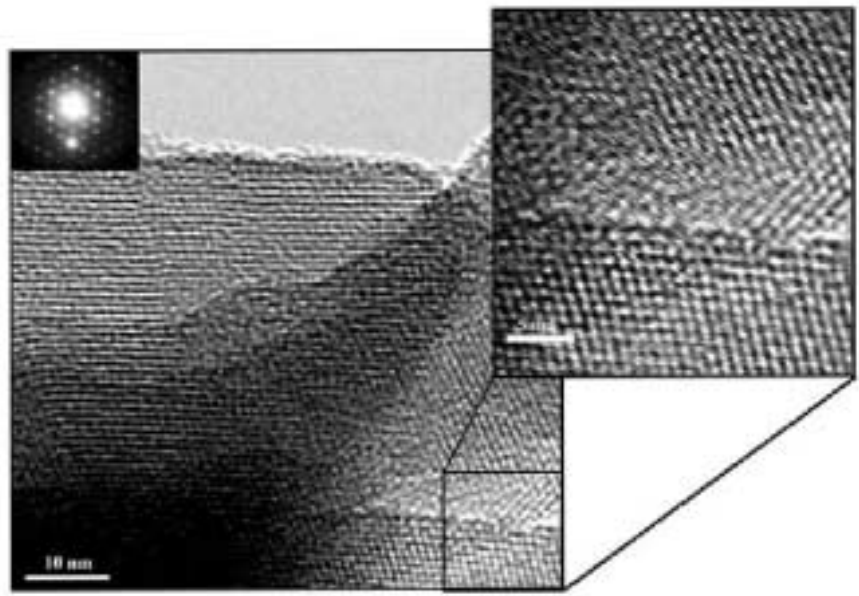

**Fig. 43.** TEM images of the ferromagnetic phase $C_{60}$ illustrating its crystalline polymeric structure [161].

Another hypothesis for the origin of unpaired spins requires closer examination of the polymerization mechanism. The disagreement between the theories and the experiment can be explained in different ways. First, the covalent bonds are not truly $sp^3$ hybridized, as the bond angles formed during polymerization are 90° rather than 109°47' as in the case of diamond. The formation energy of the interfullerene bond is smaller than that for diamond. We can assume that at high temperature the interfullerene bonds break before than the fullerene molecule is destroyed. Magnetic properties can be caused by the temperature-induced defects in the polymerization network which create unpaired spins at the interfullerene bonds. In this case analogues can be found with the disordered magnetism of nanographite, which appears in the vicinity of a heat-treatment induced insulator-metal transition, or with peculiar localized states that appear at zigzag edge of graphite ribbons.

Introducing heptagonal or octagonal rings into fullerene network produces a Gaussian negative curvature, making it possible to form doughnut-, coil- and sponge-shaped networks of carbon atoms which modify the electronic structure. The presence of a certain type of edge in fullerene network derives critical localized edge states at the Fermi level [187].

There exists a possibility that some exotic structures are formed under high pressure – high temperature treatment. It is shown that, together with polymerization, another type of chemical interaction of the molecules, called polycondensation, which leads to the formation of polymolecular structures with a shortest intermolecular distance of 0.65 nm, is possible in the system [188]. The coalescence of two molecules will inevitably result in the appearance of negative curvature. A similar defect in the graphite is called an azupyrene defect: a $\pi / 2$ local bond rotation in a graphitic network creates two heptagons and two pentagons instead of four hexagons [117]. Introduction of this defect in graphite modifies the band structure, some of which have flat excitation bands, which give rise to itinerant ferromagnetism.

The observation of ferromagnetism in hydrogenated (5.3) and oxidized (5.4) fullerenes suggests an idea that intercalated gases play significant role in magnetic properties of pressure-polymerized fullerenes. We have experimentally checked this supposition. Pressure cells were divided into two parts (Fig. 44), and one part was filled with pristine $C_{60}$, while another part contained $C_{60}$ from the same ampoule, but stored in oxygen under bright light for 24 hours. Several experiments unambiguously demonstrate that the presence of oxygen adversely affects the magnetic properties of pressure-polymerized fullerenes.

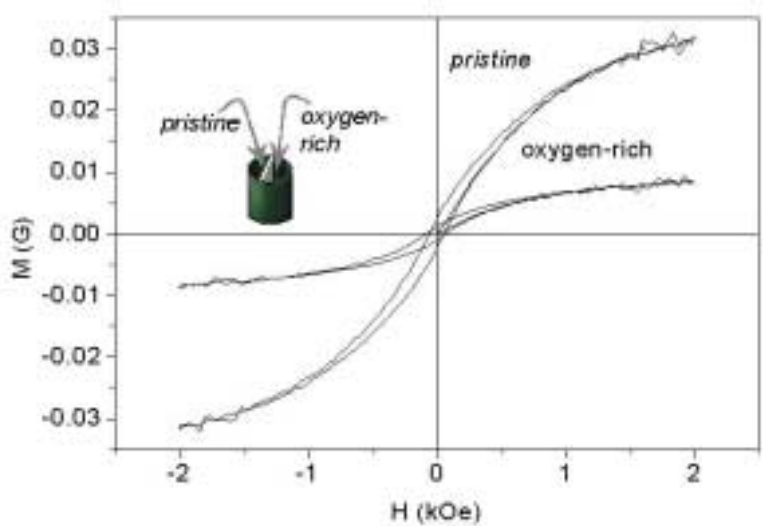

**Fig. 44 Magnetization curves for tetragonal $C_{60}$ polymers prepared from oxygen-free and from light-and-air exposed $C_{60}$ during 24 hours**

R. Wood [161] arrives to a similar conclusion: comparing the paramagnetic response of the samples, he assumes that oxygen may make a contribution, but a significant fraction of the paramagnetic signal is intrinsic to the samples and is due to the carbon radical formation. S. J. Blundell [189] points that it is difficult to construct the models of magnetism of polymerized fullerenes because the magnetization is very small and non-uniformly distributed throughout the sample; nevertheless, performing careful structural studies of samples prepared at different temperatures and pressures is the right way forward to understand this effect.

The structure and the properties of the HPHT treated fullerenes are governed by the preparation conditions as shown on the map in Fig. 45. An orthorhombic phase forms at the pressures 1 - 9 GPa and low temperatures (below 650 K). Higher temperatures give an increase in the number of interfullerene bonds, and two types of two-dimensional polymers can be formed. The tetragonal phase requires pressures about 2 GPa, whilst the optimal pressure for the rhombohedral phase is 6 GPa. This phase is more dense than the tetragonal one, and polymerization occurs in the close-packed <111> plane. At temperatures above 1000 - 1100 K the fullerene cages break down, and a new phase appears, usually described as a disordered cross-linked layered structure, disordered crystalline carbon or partially graphitized fullerene. Ferromagnetic behavior at and above room temperature in pressure-polymerized $C_{60}$ was observed for samples polymerized at 2.5 GPa, 6 GPa and at 9 GPa. For each pressure, samples with a ferromagnetic behavior could only be prepared in a particular temperature range.

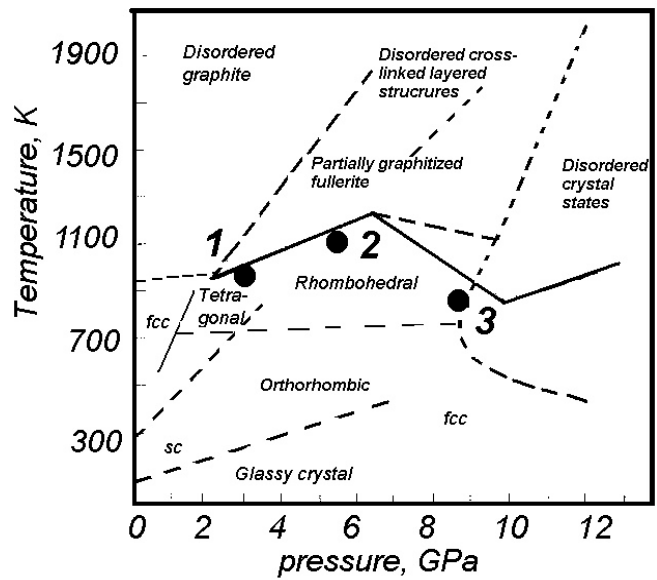

Fig. 45. A map of the pressure and temperature conditions for creating different $C_{60}$ phases (adapted from [162]). Points mark the conditions under which the ferromagnetic phase have been observed: 1 – [193]; 2 – [160], 3 – [161].

We draw attention to the fact that no one-to-one correspondence exists between the sample properties and the conditions of sample preparation. In other words, the values of the sample preparation temperature can be considered only as nominal ones. There are three reasons for this statement. First, under the high pressure conditions there are difficulties in measuring the *true* temperature, and the samples from different series, being produced at nominally identical conditions, can differ by ± 25 K in the preparation temperature. Second, an essential parameter is the treatment time: Samples prepared at the same temperature but for longer time behave as overheated samples. The third reason is the most important. The products of high-pressure treatment are inevitably inhomogeneous due to the non-uniform distribution of pressure and temperature, and the resulting patterns represent a mixture of polymerized states in the centre of the sample and graphitic states at the sample periphery. This fact significantly complicates the investigation of the electrical and magnetic properties of the individual polymerized phases.

    Carbon belongs to the class of materials with strong electronic correlations, and electronic instabilities in pure carbon can lead to various types of ordering including ferromagnetic and superconducting. Carbon nanostructures with unusually large paramagnetic moments have been discovered in a recent theoretical study: carbon nanotori, i.e. a ring of a carbon nanotube exhibits colossal paramagnetic moment [190]. Not only carbon, but hetero-nanotubes and hetero-sheets consisting of boron and nitrogen were shown to be nano-scale ferromagnets [191]. Open silicon structures are the object of molecular magnetism [192]. Ground state triplet configurations for these structures turned out to be favourable. The multiplicity of experimental data indicates an intrinsic origin magnetism of carbon-based structures. However, the nature is still controversial.